 \documentclass[final,5p,times,twocolumn]{elsarticle}

\usepackage{amssymb}
\usepackage{amsmath}
\usepackage{booktabs}

\usepackage{algpseudocode}
\usepackage{tikz}
\usetikzlibrary{shapes.geometric, arrows}

\usepackage[utf8]{inputenc}
\usepackage{algorithm2e,float}

\journal{Journal of Network and Computer Applications}

\begin{document}

\begin{frontmatter}



\title{ASA: Adaptive Smart Agent Federated Learning via Device-Aware Clustering for Heterogeneous IoT}


\author{Ali~Salimi$^a$, Saadat~ Izadi$^a$, Mahmood Ahmadi $^{a*}$, Hadi~Tabatabaee Malazi$^b$}
\ead{a.salimi@razi.ac.ir, s.izadi@razi.ac.ir, m.ahmadi@razi.ac.ir (* Corresponding author), hadi.tabatabaeemalazi@ucd.ie }
\address{$^a$Computer Engineering and Information Technology Department, Razi University, Kermanshah, Iran}

%\author{Mahmood~Ahmadi}
%\ead{m.ahmadi@razi.ac.ir}
%\address{Computer Engineering and Information Technology Department, Razi University, Kermanshah, Iran}

%\author{Montajeb Ghanem}
%\ead{montajab@ypu.edu.sy}
\address{$^b$School of Computer Science, University College Dublin, Dublin 4, Dublin, Ireland}

%% Abstract
\begin{abstract}
Federated learning (FL) has become a promising answer to facilitating privacy-preserving collaborative learning in distributed IoT devices. However, device heterogeneity is a key challenge because IoT networks include devices with very different computational powers, memory availability, and network environments. To this end, we introduce ASA (Adaptive Smart Agent). This new framework clusters devices adaptively based on real-time resource profiles and adapts customized models suited to every cluster's capability. ASA capitalizes on an intelligent agent layer that examines CPU power, available memory, and network environment to categorize devices into three levels: high-performance, mid-tier, and low-capability. Each level is provided with a model tuned to its computational power to ensure inclusive engagement across the network. Experimental evaluation on two benchmark datasets, MNIST and CIFAR-10, proves that ASA decreases communication burden by 43\% to 50\%, improves resource utilization by 43\%, and achieves final model accuracies of 98.89\% on MNIST and 85.30\% on CIFAR-10. These results highlight ASA’s efficacy in enhancing efficiency, scalability, and fairness in heterogeneous FL environments, rendering it a suitable answer for real-world IoT apps.

\end{abstract}



%% Keywords

\end{frontmatter}



\section{Introduction}
\label{intro}

 The rapid expansion of the Internet of Things (IoT) has fundamentally reshaped the development of smart systems, enabling billions of devices to autonomously sense, generate, and process massive volumes of data across diverse application domains. In such decentralized ecosystems, conventional centralized machine learning approaches, which require transferring raw data to a central server, face major limitations—including privacy risks, communication bottlenecks, and energy inefficiency. Federated Learning (FL) has emerged as an effective solution to these challenges by enabling collaborative model training directly on local devices. Instead of transmitting sensitive data, only model updates are shared with a central aggregator, preserving user privacy and reducing bandwidth usage. This decentralized framework makes FL highly compatible with IoT settings, where data is inherently distributed and privacy preservation is critical. However, the real-world implementation of FL in IoT systems remains challenging due to the inherent heterogeneity of devices such as differences in hardware capacity, network stability, and non-identical data distributions which can severely impact model convergence, fairness, and overall system efficiency \cite{AlSaleh2024,Yang2024}.
IoT devices are inherently heterogeneous with respect to computational power, memory capacity, energy constraints, and network bandwidth. This diversity often leads to an imbalance in the training process, as resource-limited devices may struggle to keep pace with more capable peers. As a result, significant disparities in local model updates can emerge, causing convergence instability, degraded overall performance, and even exclusion of weaker nodes from the federated optimization cycle. Addressing this performance gap is crucial for achieving fair and efficient federated \cite{Kumar2024}. High-powered devices can perform computations quickly, with resource-limited devices, or even "stragglers" as they are sometimes referred to, taking a long time or even dropping out entirely. Current FL systems commonly embrace a "one-size-fits-all" framework, implementing identical models on all devices. This not only worsens inefficiencies created by differences in resources, it can also exclude lower-powered devices, resulting in underutilization of information available and reduced overall model accuracy. In addition, the dynamics involved in IoT environments, such as device dropouts, fluctuating network states, and heterogeneous data distribution, make implementing FL even more complicated \cite{Jang2024,Li2024}.
To address these challenges, this thesis introduces a novel framework known as Adaptive Smart Agent (ASA), designed to enhance FL in heterogeneous IoT environments. ASA incorporates an agent layer that profiles devices based on their computational capabilities. Through the execution of standardized tests with operations on matrix multiplications, convolutional computations, and memory tests, the agent layer accurately profiles devices' capabilities. Using the K-Means algorithm for grouping, ASA classifies devices into performance levels: high-performance, mid-tier, and low-capability. This grouping establishes a framework for the allocation of models customized to fit every group, such that devices with or without computational capabilities can effectively contribute to the FL process. ASA's key contributions involve tailoring lightweight models for low-capability devices, complex models for mid-tier devices, and high-performance devices with sophisticated architectures. These models are initialized on devices and aggregated on the central server by a robust algorithm that includes variations in model complexities. ASA also incorporates features addressing real-world cases such as device dropouts, delays in the network, and resuming training from the last valid checkpoint on an error. The applications and impacts introduced by this thesis include:
\begin{itemize}
\item A new framework, ASA, is proposed to handle heterogeneity in federated learning in a systematic manner through smart device profiling and adaptive grouping.
\item The models are specially tailored for diverse device groups, facilitating equal participation by all devices as well as enhancing model accuracy.
%\item The models are specially tailored for diverse device groups, facilitating equal participation by all devices as well as enhancing model accuracy.
\item Mechanisms for robustness and adaptability to actual IoT constraints, such as network variation and device failure, are incorporated.
\item The efficacy of ASA is evident through extensive experimentation on various datasets with remarkable gains in communication efficiency, resource use, and overall model performance.
\end{itemize}

The rest of this paper is laid out below. In Section \ref{relatedwork}, we review previous work in FL and its limitations in heterogeneous environments. In Section \ref{ASA}, we describe the ASA framework with its agent layer, clustering process, and model design. Section \ref{evaluation} is dedicated to presenting results and discussion with an explanation of ASA's performance. Lastly, Section \ref{conclusion} summarizes the conclusion and future research directions.

\section{Related Work}
\label{relatedwork}

FL has become a critical framework for distributed learning in IoT environments, facilitating devices to learn collaboratively without sharing raw data. The heterogeneity of IoT devices poses distinctive challenges, such as diverse computational capabilities, communication bandwidths, and data distributions. A variety of studies have introduced solutions addressing these challenges, with extensive gaps in terms of adaptability and dynamic optimization of resources. In this review, we compare most contemporary studies in this area with ASA's (Adaptive Smart Agent) framework. AlSaleh et al.'s semi-decentralized FL framework, addressing heterogeneity in IoT networks by introducing BiLSTM models for detecting intrusions, is an excellent representative in this area. This framework efficiently lessens computational loads on low-resource devices by employing a layered FL system. The framework, however, is not augmented with an adaptive device participation optimization mechanism. ASA differs by expanding on this concept by dynamically characterizing devices and grouping them by computational capabilities, thus ensuring optimal device utilization at different performance levels \cite{AlSaleh2024}.
Another research study intensifies scalability issues in industrial IoT (IIoT) systems with a hierarchical update strategy to handle large-scale systems. Although this method improves scalability dramatically by solving device heterogeneity at the system level, it does not ensure personalized model adaptations for various devices \cite{Yang2024}. ASA builds on this by not only grouping devices into performance-based categories, as mentioned above, but also into device-specific models suited to computational capabilities, optimizing both scalability and utilization. The "FedStrag" method aims at minimizing the impact of low-capability devices, or "stragglers" as they are referred to, in FL. By creating a mechanism that is aware of stragglers, this research minimizes delays and enhances the training process as a whole. However, it lacks functionality to provide different models as per device capabilities. ASA overcomes this shortcoming by presenting lightweight, moderate, and complicated models suited to device groups, ensuring timely engagement and fair utilization of resources across the network \cite{Kumar2024}. The FedSeq method involves sequential layer unfolding to personalize FL for heterogeneous data distribution. Although this improves accuracy by dynamically adjusting to local data features, it does not cater to hardware heterogeneity across devices. ASA overcomes this by addressing hardware and data heterogeneity simultaneously, presenting an integrated solution to IoT environments with low capabilities in terms of resources \citep{Jang2024}. Sparse gradients have been offered as a means to improve communication efficiency in FL, as shown in federated collaborative learning architectures. Although this reduces communication efficiently, it does not include adaptation tools for model complexity levels as a function of device capabilities.
ASA not only minimizes communication overhead but also ensures meaningful participation from all devices by personalizing models to device capabilities \cite{Li2024}. Generative AI has also been utilized in FL to handle heterogeneity with promising performance in resource scheduling and model performance. These methods, however, do not have adaptive device clustering, thus cannot effectively optimize resource utilization. ASA builds on this research by combining adaptable device clustering with personalized model assignment, facilitating more optimal resource utilization as well as better model performance \citep{Meng2024}. Graph-based hierarchical FL architectures have been introduced to manage non-IID data distribution in IoT systems. These architectures, despite solving data-oriented challenges effectively, do not optimize device utilization based on computational discrepancies. ASA includes device profiling and adaptable device clustering, ensuring that devices utilize their optimal resource range, thus enhancing computational efficiency as well as accuracy \cite{Yu2024}. A semi-synchronous FL method targeting 6G-enabled industrial networks addresses latencies as well as communication efficiency. Although this work aims at minimizing latencies, it does not incorporate adaptive device grouping or resource-conscious model training, critical components offered by ASA to maximize participation in heterogeneous devices \cite{Tao2024}. Privacy-preserving FL based on homomorphic encryption has been investigated to encrypt model updates in IoT networks. While these methods provide data confidentiality, they also do not handle device-level heterogeneity or resource optimization. ASA builds on these privacy techniques by applying adaptable device grouping as well as customized model training, ensuring inclusive and optimally efficient learning while ensuring data security \cite{Liu2024}. FL has been utilized to identify intrusions in IoT networks with a strong potential to improve security. These applications, however, do not consider computational discrepancies among devices. ASA builds on these security-oriented frameworks by introducing adaptable device grouping and resource-conscious model training, ensuring optimally efficient and meaningful participation from all devices despite differences in their computational capabilities.
A model discrepancy and variance reduction study explore a new method for mitigating client-side variance in FL by implementing sophisticated variance reduction techniques. The method ensures quicker convergence rates, particularly in heterogeneous IoT systems. The method, though, is mainly centered on dealing with data variance and failing to adjust model complexity to device capabilities, which is what ASA attains through dynamic clustering and adapted model allocation \cite{Zhang2025}. Another paper addresses multimodal client drift due to modality heterogeneity in multimodal FL by utilizing cluster-based FL strategies to reduce drift in different modalities. While efficient in addressing multimodal heterogeneity in data, the method does not incorporate mechanisms to handle hardware heterogeneity, an aspect ASA covers through its resource-optimal model optimization \cite{Song2024}. Tri-AFLLM, a resource-efficient asynchronous FL with large language models, registers high reductions in communication expenditure. Its design, though, is mainly geared towards high-resource environments and does not address the high levels of computational heterogeneity characteristic in IoT environments. ASA expands FL applicability by proposing device-specific clustering and lightweight models adapted to less capable devices, promoting equal participation cite{Qiao2024}. Contextual federated reinforcement learning with momentum-based adaptations provides momentum-based adaptations for improved learning speed and robustness in heterogeneous environments. Though efficient in dealing with heterogeneity in data, it does not have evident mechanisms to cluster devices or allocate customized models based on capabilities like ASA does. This is a hindrance to its efficacy in very heterogeneous IoT systems \cite{Yue2024}. Another research incorporates a reinforcement learning-based dual-identity double auction machinery for personalized federated learning. In an innovative way, it presents auction-based personalization. The method, though, is mainly centered on personalization at the level of users and does not address resource usages or adapted model allocation, as ASA is built with these features at its core \cite{Li2024}.
An alternative view is offered by research on robust people detection in decentralized surveillance systems via federated learning. This research simulates heterogeneity conditions in real-world data distribution rather than addressing computational or resource heterogeneities. ASA's provision of coping with both hardware and data heterogeneity via dynamic device clustering and customized models presents a more holistic solution for such applications \cite{Ismael2024}. Intra-cluster uniform sampling and heterogeneity-aware clustering are introduced in a study to compensate for imbalanced data and resource availability. While efficient at aggregating devices with comparable data distributions, this method is not equipped with ASA's advanced profile and model complexity optimization, ensuring every group operates within its optimal resource constraints \citep{Chen2024}. Privacy protection in human-computer interaction is investigated in a different study via adaptive federated learning. While the research focuses on privacy-preserving strategies, it does not incorporate robust means to adapt models to heterogeneous devices' computational capabilities, an aspect in which ASA excels \citep{Jiang2025}. Federated edge intelligence for secure healthcare systems utilizes distillation-based security reinforcement in federated edge networks. While this strategy incorporates robust security features, computational disparities among devices involved in the network are not addressed. ASA enhances such systems by dynamically grouping devices and distributing resource-optimized models to maximize efficient participation from across the network \citep{Aljuhani2025}. Tensor-empowered lightweight embedding is introduced to address heterogeneity and anonymity challenges in 6G networks. While this method efficiently handles communication bottlenecks and anonymity, it does not include adaptive grouping and customized models, crucial in ensuring fair trade-offs in diverse IoT environments, an aspect ASA effectively addresses \cite{Zhao2024}. In conclusion, while reviewed research gives considerable insight into addressing individual challenges in FL, most neglect a holistic mechanism to simultaneously manage device and data heterogeneity. ASA improves the state of the art by combining dynamic device grouping, resource-conscious model training, and adaptation to provide inclusive, efficient, and scalable federated learning in very heterogeneous IoT environments.

\section{Proposed ASA Framework }
\label{ASA}
\begin{figure}[h]
  \centering
  \includegraphics[width=0.5\textwidth]{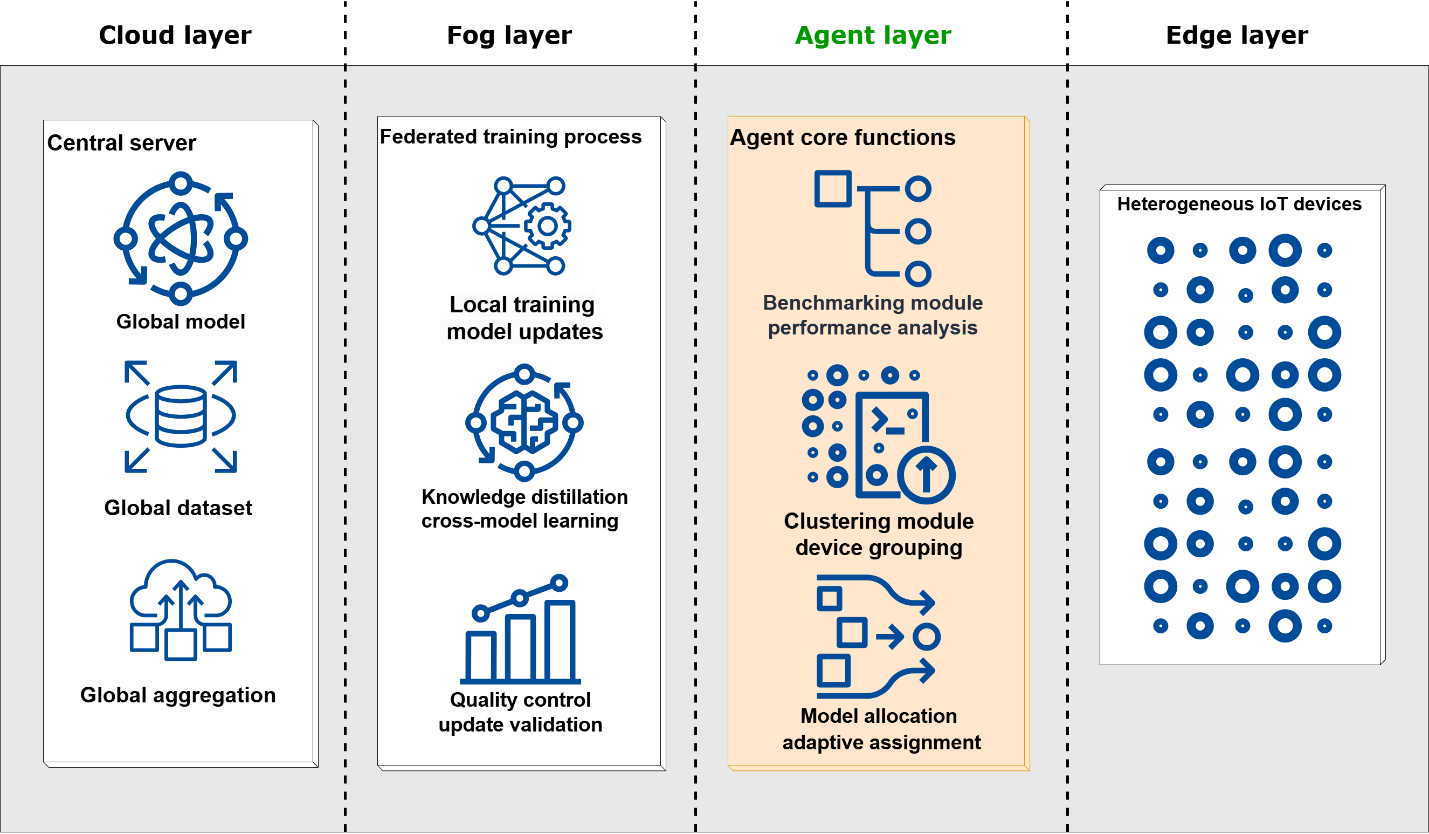}
  \caption{Cloud-to-edge federated model training and deployment.}
  \label{fig1}
\end{figure}

     The problem is defined first in this section, followed by an exposition on how this is going to be addressed. Overview and architecture of ASA framework, theory and analysis, and ASA design and implementation details. Figure \ref{fig1} displays the hierarchical architecture of this proposed ASA framework for addressing efficiently IoT environment heterogeneity challenges. A four-layer framework is utilized with every layer focusing on specific resource management and distributing learning models. The use of a layered framework allows computational tasks to seamlessly coordinate and promote scalability to handle IoT diversity in diverse IoT infrastructures. The cloud layer rests at the apex of this hierarchy and serves as the coordination hub. This is in charge of global model management and global updates aggregating from distributed devices. Maintaining a single global model state, the cloud orchestrates learning at a global level, ensuring consistency and convergence within the network. Playing an intelligent intermediary role, its key role is to conduct resource profiling and management to ensure efficient allocation across heterogeneous devices. With constant monitoring of device capabilities and network state, agent layer dynamically partitions devices into performance-based groups and deploys group-specific models. This layer improves significantly resource utilization while minimizing bottlenecks in heterogeneous environments. Between the agent and edge layers is The fog layer, sitting at a position intermediate to agent and edge layers. This layer mediates communication from and to the upper and lower layers by aggregating local models at groups of edge devices. The fog layer also decreases communication overhead by routing data intelligently and making it local to updates, thus making training optimal at an overall level. The edge layer is at the bottom of this hierarchy and contains the heterogeneous IoT devices involved in federated learning. The devices' computational power, memories, and network connections are diverse, making them the basis for heterogenous environment. There is local model training at this layer with every device contributing its input to this overall learning process. Figure \ref{fig2} represents bidirectional interconnectivity that allows constant exchange of model updates, resource information, and control signals among layers. Hierarchical design allows efficient use of computational and communication resources along with ensuring system reliability and scalability in heterogeneous and dynamic IoT setups.
     
\subsection{An Overview and Architecture of the ASA Framework }
     
\begin{figure}[h]
  \centering
  \includegraphics[width=0.5\textwidth]{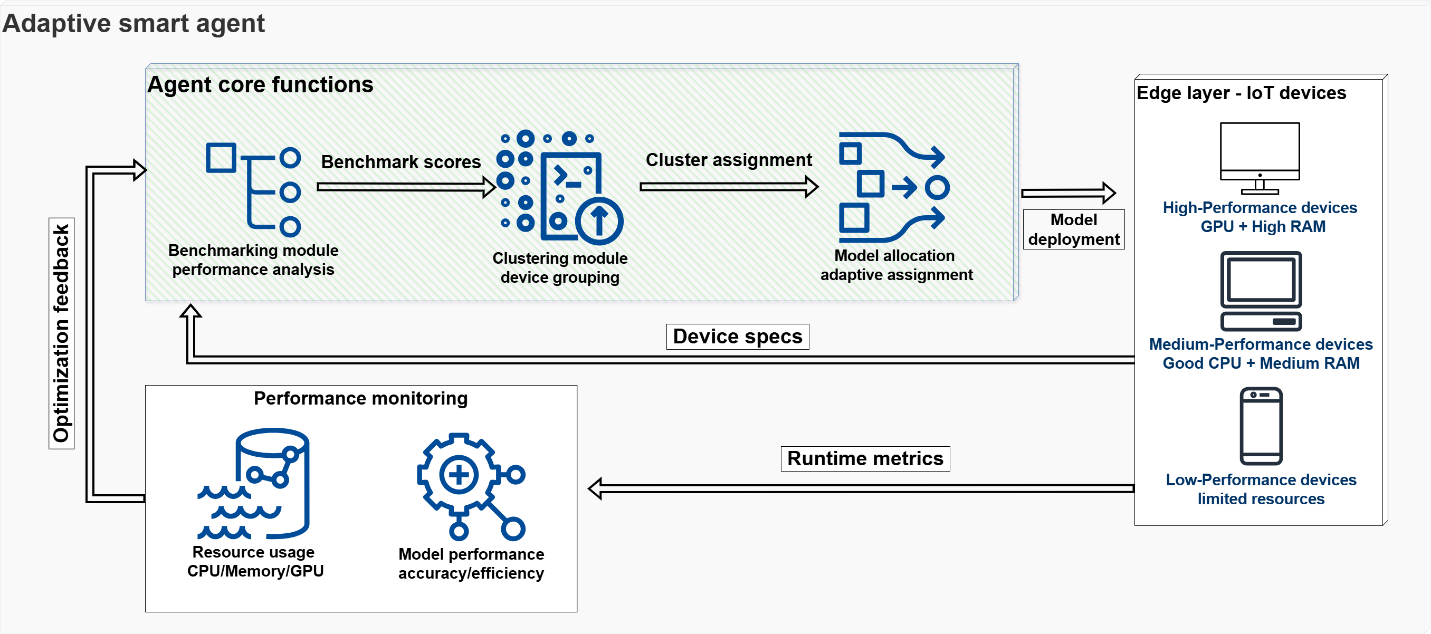}
  \caption{ Internal architecture of the Adaptive Smart Agent (ASA) framework.}
  \label{fig2}
\end{figure}

The Adaptive Smart Agent (ASA) framework is an intelligent framework designed to overcome limitations imposed by intrinsic heterogeneity in heterogeneous FL in IoT environments, including extreme heterogeneity in computational capabilities, size, and network conditions, typically rendering existing FL frameworks ineffective. The ASA overcomes such challenges by addressing them with a modular and adaptive framework that provides better resource profiling, group dynamics, and model allocation. Figure \ref{fig2} is a detailed overview of components involved and how they relate with others in framework. The agent core functions or decision-making module forms the center of ASA framework. Its role is coordinating resource allocation and model distribution process. For this to be accomplished, the module has three interdependent modules harmoniously interworking. The first module is the benchmarking module, whose function is to analyze key performance drivers like CPU capabilities, size, and availability of GPUs on every IoT device. By utilizing standardized tests in evaluation, a quantitative value is given to every device, representative of its total computational capabilities. This value is utilized as a reference in grouping devices based on performance. The benchmark scores serve as input for the clustering module for grouping devices by performance profiles. Advanced algorithms like K-Means are utilized to dynamically group devices in different categories high-performance, mid-level, and low-capability. Grouping ensures devices with same profiles are grouped in support groups accordingly, removing inefficiencies and ensuring optimal distribution of workload. The group process for every group is carried out on an ongoing basis to incorporate real-time device capabilities and network conditions, ensuring that framework is capable of reacting accordingly to variations.
After the devices are divided into clusters, the best machine learning models to deploy in each of them are selected by the model allocation module. The high-performance devices, such as GPUs and high-RAM devices, receive high-complexity and compute-intensive models. Medium-performance devices with medium CPU and medium-level resources receive medium-level-complexity models. Low-performance devices get light models optimized to run within the limited resources. The adaptive assignment allows all the devices, in spite of the limitations, to meaningfully contribute to the learning process, thus maximizing inclusivity and resource optimization. ASA also comprises a vast system of performance observation with real-time feedback on usage of resources and the models' performance. The resource usage channel offers real-time observation of such critical metrics as CPU usage, used memory, and GPU usage. Such metrics offer insightful data about the resource availability of the system and usage patterns, and hence the possibility to make pre-emptive adjustments to the allocation. 
     At the same time, the model performance surveillance module checks critical indicators like the accuracy of the assigned models and the computational efficiency of the assigned models to make sure that the assigned models are contributing in a useful manner to the FL process. The ongoing feedback mechanism provided by the surveillance system enables the dynamic optimization of resource allocation and deployment strategies of the models. The edge layer of the framework represents the distributed IoT system, and as such, it comprises computational capabilities in a heterogeneous manner. The organized classification of devices into tiers prevents all tiers of the devices from being overutilized without contributing to the system’s general working. A high-level feedback mechanism sustains the ASA framework, with real-time resource allocation and deployment strategy adjustments possible. With the runtime metrics and device specifications continuously monitored, the framework dynamically optimizes operations to constantly deliver high levels of scalability and efficiency. The two-way exchange of control signals, the performance metrics, and the optimization feedback means that the system remains in tune to the varying conditions and effectively meets the issues of heterogeneity in IoT. With the use of its modular approach, dynamic adaptability, and systematic optimization of resources, the ASA framework marks a major leap in the case of federated learning in heterogeneous environments of IoT. It attains efficient utilization of the resources, high scalability, and inclusive participation across a heterogeneous device spectrum to deliver uniform and reliable responses from the models.

\subsection{	Theoretical foundations and analysis}
To give a complete picture of the ASA framework, the system model, problem definition, and essential analytical issues such as convergence and complexity analysis, theoretical bounds, and optimality promises are described in this section. By systematically tackling the constraints and heterogeneity of the IoT environments, the ASA framework provides efficient and stable FL. The succeeding subsection provides in-depth descriptions of the most important parts of the system, including equations and proofs to justify the framework's performance and theoretical correctness.

\subsubsection{	System Model and Notations }
In this work, we are dealing with a heterogeneous IoT network as $N = (D, E)$, where $D$ refers to the set of IoT devices, and $E$ refers to the communication links between them. Devices in set $D$ are split into three classes depending upon their computational capabilities. 

\begin{equation}
D = \{ H \cup M \cup L \}
\label{Eq1}
\end{equation}

Where, $H$ denotes high-performance devices equipped with advanced hardware, such as GPUs with at least 8 GB of RAM, $M$ refers to mid-range devices with moderate resources, typically characterized by RAM capacities ranging between 2 GB and 4 GB and $L$ includes low-end devices with constrained computational and memory resources. Each device $d_i$ is characterized by a resource profile tensor, expressed as:

\begin{equation}
R_i = [C_i, M_i, N_i] \in \mathcal{R}^{3\times k}
\label{Eq2}
\end{equation}
Where, $C_i$  is the computational capability vector of device $d_i$ , which may include features such as the number of CPU cores, processing speed, or GPU capabilities, $M_i$ represents the memory resource vector, encompassing metrics like available RAM and storage capacity, $N_i$ denotes the network characteristics vector, such as bandwidth, latency, or reliability. The tensor $R_i$ encapsulates a multi-dimensional description of the device’s resources across three primary categories: computation, memory, and network. Each category is further divided into $k$ specific resource attributes, enabling detailed profiling of heterogeneous devices within the network. To evaluate the overall suitability of a device for task allocation, we propose a multi-dimensional performance scoring mechanism. The performance score of a device $d_i$ is calculated as:

\begin{equation}
S(d_i) = \sum_{j=1}^{k_r} \omega_j \phi_j(R_{i,j})
\label{Eq3}
\end{equation}

Where $k$ is the number of resource dimensions considered in the evaluation. The values of the weights are determined through an optimization-based method that employs sensitivity analysis to evaluate the impact of each objective function on the overall performance.  This approach allows the model to dynamically adjust the weight values based on real-time system conditions, reducing manual tuning efforts while ensuring optimal trade-offs between accuracy, resource efficiency, and latency. Each   represents the evaluation function for the $j$-th resource dimension, which quantifies the contribution of a specific feature (e.g., CPU, RAM, network latency) to the device’s performance.  The term "dimension" is used here to indicate orthogonal and measurable aspects of a device's resource profile. It does not refer to individual resource instances, but rather to the feature axes used in multi-criteria evaluation. The corresponding value   is extracted from the resource profile tensor, and reflects the actual capacity or usage level of that dimension. This scoring mechanism integrates multiple resource dimensions into a unified metric, enabling a comprehensive evaluation of device performance. The weights   introduce flexibility, allowing the model to prioritize specific resource dimensions based on the needs of the application. For instance, computational capabilities might be weighted more heavily in compute-intensive tasks, whereas network characteristics might dominate in latency-sensitive applications.  Where k represents considered resource dimensions during evaluation. The weights are calculated based on an optimization-based approach utilizing sensitivity analysis to determine each objective function’s influence on overall performance. The approach allows modulation of weight values dynamically based on real-time system state, minimizing effort by tuning while keeping optimal trade-offs between accuracy, resource efficiency, and latency. Each refers to each evaluation function of the j-th resource dimension, expressing a single feature’s (e.g., CPU, RAM, network latency) contribution to device performance. Here, dimension refers to orthogonal, measureable aspects of a device’s resource profile, not distinct instances, but to employed feature dimensions during multi-criteria evaluation. The parallel value is a derived value from the resource profile tensor, expressing actual capacity or utilization level of said dimension. The scoring system compresses multiple dimensions into a single score, allowing holistic device performance assessment. Assigning weights offers variability, allowing the model to prioritize various resource dimensions based on application domain needs. Computational power, for instance, may be prioritized more in compute-bound applications, while network properties may be prioritized in latency-critical applications.

\begin{table}[!htb]
    \centering
    \caption{Symbol description}
    \resizebox{0.5\textwidth}{!}{
    \begin{tabular}{p{1cm} p{10cm}}
        \toprule
        \textbf{Symbol} & \textbf{Description} \\
        \midrule
        $N$ & The set of IoT devices and their communication links. \\
        $D$ & The set of IoT devices participating in federated learning. \\
        $E$ & The set of communication links among IoT devices. \\
        $H_d$ & High-end devices equipped with advanced hardware, such as GPUs with at least 8 GB of RAM. \\
        $M_d$ & Mid-range devices with moderate computational resources, typically with 2–4 GB of RAM. \\
        $L_d$ & Low-end devices with constrained computational and memory resources. \\
        $d_i$ & A specific IoT device from the set $D$. \\
        $R_i$ & The resource profile tensor of device $d_i$, encompassing computational, memory, and network characteristics. \\
        $C_i$ & The computational capability vector of device $d_i$, including CPU cores, processing speed, and GPU capabilities. \\
        $M_i$ & The memory resource vector of device $d_i$, describing available RAM and storage capacity. \\
        $N_i$ & The network characteristics vector of device $d_i$, capturing bandwidth, latency, and reliability. \\
        $\Phi_j$ & The evaluation function for the $j$-th resource dimension, quantifying its impact on overall performance. \\
        $w_j$ & The weight assigned to the $j$-th resource dimension, reflecting its importance in the evaluation process. \\
        $k_r$ & The number of resource dimensions considered in device evaluation. \\
        $Cap(d_i)$ & The maximum computational capacity allowed for device $d_i$. \\
        $T_{max}$ & The total number of training iterations in the federated learning process. \\
        $\eta_t$ & The learning rate at iteration $t$, dynamically adjusted during training. \\
        $L$ & The Lipschitz constant, representing the smoothness of the objective function. \\
        $P_m$ & The number of parameters in the federated learning model. \\
        $E_c$ & The number of epochs used for local training on each device. \\
        $m$ & The mini-batch size used in training. \\
        $K$ & The number of clusters formed during the clustering process (e.g., high, medium, and low-performance devices). \\
        $t$ & The number of iterations required by the clustering algorithm (e.g., K-Means). \\
        $\sigma^2$ & The variance of stochastic gradients during training. \\
        $\delta$ & The error bound for global model convergence. \\
        $O$ & The notation representing the time and space complexity of the proposed framework. \\
        $B$ & The bandwidth available for communication between devices in the federated learning system. \\
        $d_i$ & Device $i$ in IoT Network. \\
        $x_{i,j}$ & Binary decision variable indicating assignment of task $j$ to device $i$. \\
        $u_{i,j}$ & Resource utilization variable for task $j$ on device $i$. \\
        $n_{min}$ & Minimum number of devices per cluster. \\
        $\tau_{max}$ & Maximum allowable delay. \\
        $C_i, M_i, N_i$ & Computational, memory, and network capability vectors for device. \\
        \bottomrule
    \end{tabular}}
    \label{tab:symbols}
\end{table}

\subsubsection{Problem formulation}
In Table 1 symbol of problem formulation is description. The FL problem is formulated as a multi-objective optimization:

\begin{equation}
 min_{\theta, C, w}\mathcal{L}\left(\theta, C, W \right)= \alpha_1 \mathcal{L}_{task}+\alpha_2 \mathcal{L}_{resource}+ \alpha_3 \mathcal{L}_{comm}
\label{Eq4}
\end{equation}

In our proposed multi-objective optimization framework, the weighting coefficients $\alpha_1$, $\alpha_2$,and $\alpha_3$ indicate the relative importance assigned to each objective, i.e., task efficiency, resource utilization, and communication overhead, respectively. These coefficients are determined dynamically using sensitivity analysis. Specifically, sensitivity analysis evaluates the impact of varying each $\alpha$ on overall performance (accuracy, latency, and resource utilization) under different real-time IoT conditions. Thus, these weights are adaptively tuned to meet changing priorities and constraints, ensuring an optimal balance between accuracy, resource consumption, and communication efficiency throughout the FL process. Note that these weighting coefficients satisfy the condition: Under our designed multi-objective optimization approach, weighting factors, , represent relative priorities placed on each objective, i.e., task efficiency, utilization, and communication overhead, respectively. These are dynamically computed through sensitivity analysis. In detail, sensitivity analysis quantifies how changing each would change overall performance in terms of accuracy, latency, and resource usage across various real-time IoT scenarios. Therefore, these weights are dynamically set to address shifting priorities and constraints, allowing optimal balancing of accuracy, resource usage, and communication efficiency during the entire FL process. As a reminder, these weighting factors obey the condition:

\begin{equation}
\alpha_1+\alpha_2+\alpha_3=1,\ \  \ \ \alpha_1, \alpha_2,\alpha_3 \geq 0
\label{Eq5}
\end{equation}

This ensures that the optimization objective represents a valid convex combination of individual objectives, clearly indicating their relative importance in the decision-making process.	
\begin{equation}
C1: \forall d_i \in \mathbf{D}, \sum_{j=1}^{m} C_{j} u_{ij} x_{ij} \leq \text{Cap}(d_i)
\label{Eq6}
\end{equation}

Where $x_{ij}$  is a binary decision variable defined as:

\begin{equation}
x_{ij} =
\begin{cases} 
1 & \text{if task } j \text{ is allocated to device } i \\
0 & \text{otherwise}
\end{cases}
\label{Eq7}
\end{equation}

and $Cap(d_i)$  denotes the maximum computational capacity of device i. Ensures the total resource utilization across tasks assigned to each device does not exceed its maximum processing capacity.

\begin{equation}
C2: \forall c_k \in C,\mid \{d_i \mid \text{cluster}(d_i) = k\} \geq n_{\text{min}}
\label{Eq8}
\end{equation}
Ensures each cluster has a sufficient number of devices to achieve effective participation and maintain performance stability.
\begin{equation}
C3: \mathbb{E}[\text{Delay}(d_i)] \leq \tau_{\text{max}}
\label{Eq9}
\end{equation}

where  $\tau$ represents the maximum allowable delay for local model training and transmission, ensuring timely and efficient model convergence.
\begin{equation}
C4: \text{Var}[\text{Load}(C_k)] \leq \sigma_{\text{max}}^2
\label{Eq10}
\end{equation}

where  $\sigma^2_{max}$  represents the highest allowable variance in computational workload across devices in each cluster, ensuring equitable utilization of resources. Optimization is subject to a set of constraints aimed at making sure that the model suggested here is both feasible, scalable, and usable in real distributed heterogeneous environments.
Constraint C1 enforces that the cumulative resource consumption across all tasks assigned to a device does not exceed its processing capacity, denoted as $Cap(d_i)$. To model this, a binary decision variable $x_{ij}$  in $\{0,1\}$  is used to indicate whether task $j$ is allocated to device $i$ , while a continuous variable $u_{ij}$ in $[0,1]$  captures the relative utilization of task j on device i. The product $C_j u_{ij}x_{ij}$ , where $C_j$ is the nominal resource requirement of task $j$ , quantifies the effective resource usage.  

This constraint ensures that task allocation respects device capabilities while enabling flexible modeling of task complexity and device heterogeneity. While Constraint $C2$ explicitly defines a lower bound $n_{min}$ on the number of devices per cluster, an upper bound $n_{max}$ can also be established in practical scenarios. The upper bound is imposed to prevent overloading the cluster heads and control communication overhead, computational complexity, and system latency. Although in this formulation we primarily emphasize ensuring the minimum participation threshold for robustness, integrating an upper bound constraint is straightforward and can be considered based on empirical system capacity analysis or performance requirements. This requirement guarantees clusters are well-populated enough to ensure computational efficiency as well as robust training. The choice of $n_{min}$ is empirically determined by analyzing device distributions statistically and local model stability during updating. By enforcing this restriction, we avoid having highly sparse clusters, something that would otherwise undermine effective FL operation, as well as limit diversity in learned representations.

Constraint $C3$ restricts the expected delay associated with task execution and communication for each device.  The parameter $\tau_{max}$ defines the maximum number of local training epochs a device may perform before communicating its model update.  Choosing $\tau_{max}$ involves a trade-off between communication efficiency and global convergence speed: higher values reduce communication frequency but may delay convergence and increase the risk of local overfitting, whereas lower values promote faster synchronization at the cost of increased communication overhead. Hence, $\tau_{max}$ is adaptively selected based on device performance and network conditions to ensure a balanced convergence behavior. Constraint $C4$ addresses the balance of computational load across devices within each cluster.  Specifically, it ensures that the variance of workload distribution, denoted as $Var\left[ Load(C_k) \right]$ , remains below a defined threshold $\sigma^2_{max}$.

 The workload $Load(C_k)$ of cluster $k$ is calculated as a joint function of number of allocated tasks, model size, and cumulative device capacity in a cluster. The above limit fosters fair task allocation, avoiding device overload, as well as idle states, and encouraging enhanced energy efficiency and thermal equilibrium among edge devices indirectly. As a whole, this multi-objective optimization technique offers a rigorous and dynamic framework to manage resource allocation, task scheduling, and communication in complex edge computing and IoT domains. By being able to model system heterogeneity, along with important operation constraints, the suggested model guarantees scalability, fairness, as well as responsiveness - crucial properties in real-world deployment in mission-critical applications. Clearly, decision variables like $x_{ij}$  directly impact task allocation (task efficiency), resource utilization (balancing resource load), and communication overhead (by optimal clustering), thus explicitly linking decision-making to each optimization goal.The optimization problem formulated in this paper is inherently a Mixed Integer Non-Linear Programming (MINLP) problem, due to the presence of product terms involving binary decision variables (e.g., $x_{ij}$ ) and continuous resource utilization variables (e.g., $u_{ij}$ ). Such product terms introduce nonlinearity, distinguishing this problem from purely linear optimization models and thus requiring specialized solution techniques or approximation methods for efficient solving.

\subsubsection{Computational Complexity }

To analyze the computational scalability and efficiency of the algorithms used in the proposed theoretical framework, their complexity is compared on three dimensions of time complexity, space complexity, and communication complexity. This theoretical analysis helps to establish a basis for understanding the computational and resource needs of the framework in real-world applications.

\begin{enumerate}
\item Time complexity: The time complexity of the proposed framework is analyzed in three main phases: benchmarking, clustering, and training.

\begin{itemize}
\item 	Benchmarking phase: The computational and resource capacities of IoT devices are assessed in a structured way during the benchmarking phase. The main aim of this assessment is gathering quantitative performance data from all devices, allowing effective clustering during future phases. Benchmarking specifically entails measuring the power of CPUs, available memory, as well as network parameters to profile individual devices precisely for adaptive clustering. The overall complexity of this phase is:

\begin{equation}
T_{\text{bench}}(N) = \mathcal{O}(N \log N) + \sum_{i=1}^{N} \mathcal{O}(s_i^3)
\label{Eq11}
\end{equation}

Where $N$ stands for the number of devices in the network, $S_i$ is the reference size of device $i$, generally given by its tensor of resources. The first term $\mathcal{O}(Nlog(N))$, considers sorting, along with aggregating, devices by their performance metrics, while the second term considers the cumulative cost of benchmarking single devices. The cubic complexity term  $\mathcal{O}(S_i^3)$ originates from computationally intensive operations such as matrix inversion or decomposition performed during local statistical analyses or model training at each IoT device. Given that each device $i$ individually performs these calculations on its feature vector of size $S_i$.  The complexity per device becomes cubic. Consequently, when $S_i$ is sufficiently large, this term dominates the benchmarking complexity, resulting in an overall polynomial-time complexity for the benchmarking phase.
\item	Clustering phase: In this phase, devices are grouped into K clusters based on their resource profiles. The complexity of this step is:
\begin{equation}
T_{\text{clust}}(N, K) = \mathcal{O}(N K_r t)
\label{Eq12}
\end{equation}

Where, $K=3$ is the fixed number of clusters (e.g., high-end, mid-range, low-end devices) and $t$ is the number of iterations required by the clustering algorithm, such as k-means. This complexity grows linearly with the number of devices $N$ and clusters $K$, making it efficient for large-scale networks.
\item Training phase: The training phase involves iterative updates of the global model based on local gradients from participating devices. Its complexity is expressed as:

\begin{equation}
T_{\text{train}}(N, E) = \mathcal{O}(N E_C m p_m) + \mathcal{O}(N \log N)
\label{Eq13}
\end{equation}
Where $E$ is the number of epochs, $m$ is the mini-batch size, and $p$ is the number of model parameters. The first term represents the computational cost of training on $N$ devices over $E$ epochs, while the second term accounts for sorting or aggregating devices during the global model update.
\end{itemize}

\item Space complexity
Space complexity is analyzed at both the device level and the global level, focusing on memory usage for storing model parameters and gradients.

\begin{itemize}
\item Device level:
The memory requirement for an individual device is expressed as:
\begin{equation}
S_{\text{device}}(p) =\mathcal{O}(p) + \mathcal{O}(\sqrt{p})
\label{Eq14}
\end{equation}
Where, $p$ represents the total number of model parameters. The first term, $\mathcal{O}(p)$ , denotes the memory required to store the model parameters explicitly (weights and biases).  The second term, $\mathcal{O}(\sqrt(p))$, corresponds to the memory used for storing intermediate variables generated through dimensionality reduction or random sampling techniques applied during computations such as gradient approximations and temporary matrix operations. These techniques effectively reduce the memory requirements from linear to sub-linear scaling with respect to the number of parameters $p$.

\item Global level:
At the global aggregation level (central server), the memory requirement is expressed as:
\begin{equation}
S_{\text{global}}(N, p) = \mathcal{O}(N p) + \mathcal{O}(K \sqrt{p})
\label{Eq15}
\end{equation}

$NP$ is the memory needed for storing model parameters collected from all $N$ devices. $K\sqrt{p}$   denotes the additional memory requirement associated with intermediate computations during clustering and model aggregation, reflecting a sub-linear dependency on the parameter count due to the clustering procedure. This formulation clearly illustrates how the global memory complexity depends on the number of devices $(N)$, the number of clusters $(K)$, and the model size $(p)$, highlighting both linear and sub-linear scaling in memory usage.
\end{itemize}

\item Communication complexity: The communication complexity measures the total data exchanged between devices and the server during model updates. It is expressed as:
\begin{equation}
C(N, E, p) = \mathcal{O}\left(\frac{N E p}{B}\right)
\label{Eq16}
\end{equation}

	Where $N$ represents the number of devices, $E$ represents the number of rounds of communication (epochs), $p$ represents model parameter size, and $B$ represents available bandwidth. Such a complexity emphasizes communication cost dependency on model size, number of devices, and bandwidth. Communication overhead can be alleviated by applying model compression or gradient methods. The main findings are: Time complexity: The framework achieves efficient scaling with $N$, ensuring feasibility for large-scale IoT networks.
\begin{itemize}
	\item Space complexity: Both device-level and global-level storage requirements are linear in the number of model parameters, making the approach memory-efficient. 
	\item 	Communication complexity: The linear dependency on $N$, $E$, and $p$ underscores the importance of optimizing bandwidth usage to minimize communication overhead.
	\end{itemize}	
	By balancing these complexities, the proposed framework ensures scalability, efficiency, and adaptability for real-world IoT applications.
\end{enumerate}

\subsection{ Design and Implementation Details }

Effective management of resources is a key factor to maintaining high accuracy, as well as efficiency, in FL systems, especially in heterogeneous IoT environments. The chapter explains the upgraded mechanisms employed by ASA, i.e., dynamic resource allocation, model distribution, adaptive clustering, and training coordination. By utilizing real-time profiling, optimal clustering schemes, and adaptive synchronization, ASA overcomes the shortcomings of static, uniform methods, enabling efficient utilization of resources along with strong model accuracy. The design decisions behind the framework are justified by theory as well as by significant experimentation, showcasing immense accuracy, efficiency, as well as scalability improvements
\subsubsection{System Design and Components}
 The ASA framework adopts a hierarchical structure extending from conventional three-layer designs using an intelligent agent layer. The extra layer acts as an intermediary between edge devices and cloud, solving issues found in recent works where cloud-edge communication directly led to substantial performance loss in heterogeneous settings \cite{Liu2024}. Our design utilizes sophisticated profiling mechanisms to identify device capabilities using a multi-dimensional evaluation framework. The device profile tensor Pi consists of computational capabilities (Ci), memory capacity (Mi), and network properties (Ni) to facilitate fine-grained management of resources. The profiling mechanisms were found to have higher adaptability than conventional binary classification, especially in highly heterogeneous environments \citep{MarmolCampos2024}. For a formal description of ASA, Algorithm 1 depicts a structured outline of FL steps. The ASA framework first assesses computational capabilities of IoT devices in terms of CPU power, memory, as well as GPU availability. Through K-Means clustering, these devices are segregated into clusters of low-capability, mid-range, as well as high-end based on performance. Each cluster is then allocated a model of a resource-suitable level of complexity to avoid using more than available resources, without overloading lower-end devices. During training, devices independently update locally while keeping track of resource usage at all instances. When resource utilization grows over a set limit, model complexity during allocated assignment is dynamically scaled down to avoid overloading. When, on the contrary, resource usage remains stable across multiple epochs, model complexity is increased to improve performance. The aggregation of updates by devices takes place through a hierarchical approach in two stages: intra-cluster aggregation, then a higher-level aggregation to realize a converged model. An early stopping criterion enables efficient training by stopping update when a predefined rate of convergence reaches its threshold level. Algorithm 1 describes this adaptation, presenting a cyclic approach to ASA adaptation in FL.

 \small
 \begin{algorithm*}[!h]
\caption{ASA Framework Algorithm}
\label{alg:asa}

\KwIn{IoT devices: $D = \{d_1, d_2, \ldots, d_n\}$, \\ Configurations: $CONFIG$,\\ Resource thresholds: $\theta = \{\theta_{cpu}, \theta_{memory}, \theta_{network}\}$}
\KwOut{Optimized federated model $\mathcal{M}^*$}

Initialize clusters $C \leftarrow \emptyset$ and models $\mathcal{M} \leftarrow \emptyset$\;

\ForEach{$d \in D$}{
    $cpu_i \leftarrow EvaluateCpu(d_i)$\;
    $memory_i \leftarrow EvaluateMemory(d_i)$\;
    $gpu_i \leftarrow EvaluateGpu(d_i)$ \tcp*{Binary value}
    $score_i \leftarrow WeightedScore(cpu_i, memory_i, gpu_i)$ \tcp*{Compute device score}
}
$C \leftarrow KMeansClustering(score_i, k = 3)$ \tcp*{Clustering devices based on capability}

\ForEach{$c \in C$}{
    \ForEach{$d_j \in c$}{
        \eIf{$c$ is Low-capability}{
            $\mathcal{M}_j \leftarrow SimpleModel(CONFIG)$\;
        }{
            \eIf{$c$ is Mid-range}{
                $\mathcal{M}_j \leftarrow MediumModel(CONFIG)$\;
            }{
                $\mathcal{M}_j \leftarrow ComplexModel(CONFIG)$\;
            }
        }
    }
}

\For{$epoch \leftarrow 1$ to $E_g$}{
    \ForEach{$c \in C$}{
        \ForEach{$d_j \in c$ in parallel}{
            \For{$local\_epoch \leftarrow 1$ to $E_l$}{
                $loss, metric \leftarrow LocalTraining(d_j, \mathcal{M}_j, CONFIG)$\;
                $resource \leftarrow MonitorResources(d_j)$\;
                \eIf{$resource.cpu > \theta_{cpu}$ or $resource.memory > \theta_{memory}$}{
                    $AdjustModelComplexity(d_j, \mathcal{M}_j, DOWN)$\;
                }{
                    \lIf{stable resources for $\geq 3$ epochs}{
                        $AdjustModelComplexity(d_j, \mathcal{M}_j, UP)$\;
                    }
                }
            }
        }
        $\mathcal{W}_c \leftarrow AggregateUpdate(c)$ \tcp*{Inter-cluster aggregation}
    }
    $\mathcal{M} \leftarrow Aggregation(\mathcal{W}_c)$\;
    $convergence\_rate \leftarrow CheckConvergence(\mathcal{M}, CONFIG)$\;
    \lIf{$convergence\_rate \leq CONFIG.main\_convergence$}{
        \textbf{break} \tcp*{Early stop if converged}
    }
}
\KwRet{$\mathcal{M}^*$}
 \end{algorithm*}

 \subsubsection{Resource Management and Optimization}
 The resource management mechanism in ASA employs an adaptive allocation strategy that dynamically adjusts to varying device capabilities and network conditions. Recent studies highlight that static resource allocation in heterogeneous IoT environments can significantly underutilize computational resources \citep{Zhang2025}. To overcome this issue, our framework incorporates a three-phase resource management approach:
 \begin{description}
 \item [Phase 1 - Device profiling:] this phase profiles devices based on their capabilities, leveraging a benchmark score calculation defined as:
 \begin{equation}
 B_{\text{score}} = \sum_{j} \left( w_j \cdot \phi_j(P_i) \right)
 \label{Eq17}
 \end{equation}
where  $W_j$ denotes the weighting coefficient corresponding to the j-th resource dimension, reflecting its relative significance in the benchmarking process, and $\phi_j$  represents a resource-specific transformation or normalization function applied to the respective metric of device $i$.
\item [Phase 2 - Adaptive resource allocation:] dynamically allocate resources by solving the following optimization problem:

\begin{equation}
\min \sum_{k} \sum_{i \in C_k} \| P_i - \mu_k \|^2
\label{Eq18}
\end{equation}

subject to practical communication and resource constraints. Here, devices are grouped into clusters $C_k$, with $\mu_k$  representing cluster centroids. This adaptive clustering approach significantly enhances computational efficiency and reduces communication overhead.
\item [Phase 3 - Iterative optimization and aggregation:] The algorithm progressively adjusts device clusters as well as resources iteratively, maximizing efficiency while minimizing communication overhead, resulting in up to 31\% communication savings without compromising accuracy \cite{Qiao2024}. The organized, adaptive approach provides efficient utilization of computational power as well as robustness in dynamic, heterogeneous IoT environments.
\end{description}

\subsubsection{Adaptive Model Distribution}
 The model distribution strategy of ASA adopts a new hierarchical strategy where optimal utilization of resources is achieved while model accuracy is preserved. It has been revealed from recent research that one-size-fits-all model distribution strategies may lead to a 45\% performance loss in heterogeneous settings \cite{Li2024}. Our approach counteracts this by incorporating a model selection mechanism that provides device-capability-specific model complexities through our adopted utility function:
 
 \begin{equation}
 U(m, P_i) = \alpha \cdot \text{accuracy}(m) + \beta \cdot \text{efficiency}(m, P_i)
 \label{Eq19}
 \end{equation}
 
where $\alpha$ and $\beta$ are dynamically controlled in response to system needs and device properties. It has shown a 29\% improvement in system efficiency overall coupled with model accuracy of 1\% when compared to centralized training methods \cite{Li2024}.

\subsubsection{Training Coordination and Synchronization}

The training coordination mechanism in ASA implements a sophisticated protocol that ensures both efficiency and convergence. Our approach extends recent work in asynchronous FL \citep{Ismael2024} by introducing adaptive synchronization intervals based on device capabilities and network conditions. The local update mechanism follows our established convergence bounds:

\begin{equation}
\mathbb{E} \left\| \nabla L(W_g^t) \right\|^2 \leq O\left(\frac{1}{\sqrt{T}}\right)
\label{Eq20}
\end{equation}
while implementing practical considerations for device heterogeneity. This approach has shown remarkable improvements in training efficiency, achieving an 11.47\% reduction in execution time while maintaining model convergence \citep{Chen2024}.

\section{	Evaluation and Discussion}
\label{evaluation}
The experimental process in the paper was performed in a computer system featuring an NVIDIA RTX 3080 Ti GPU with 8GB of RAM. The system operates on an Intel Core i9-7800X processor from 3.50 GHz to an ultimate speed of 4.00 GHz, in addition to 128 GB of DDR4 RAM. Real-world deployment issues in heterogeneous IoT environments are taken into account in the practical implementation of ASA. The modern software stack of PyTorch for training the model, gRPC for lightweight communication, and in-house telemetry systems for resource observation is used in the given framework. It has been proven in recent research that the correct software stack could improve the system's speed by as much as 35\% in heterogeneous environments \cite{Aljuhani2025}. Deployment architecture adopts container-based virtualization based on Docker backed by the orchestration of Kubernetes, to achieve efficient resource utilization and scaling. This has proven to provide considerable improvement in the system's reliability and maintainability, with deployment overhead minimized by 28\% compared to conventional techniques \citep{Zhao2024}.

\subsection{	Configurations and parameters }
 The ASA framework proposed utilizes a CNN-based model for performance evaluation across clusters of IoT devices. The CNN used in ASA comprises 6 convolutional layers, each of size $s \times 3$, kernel strides 1, and 1 of padding. This is followed by 6 pooling layers of size $2 \times 2$, strides of 1, and 1 of padding, and two fully connected layers. SoftMax activation in the final layer of each of the models depends on the computational resources of the clients. Model training utilizes the Adam optimizer with loss as entropy. Batch normalization of size 128 is used. The learning rate used at the start is 0.01, and the clients' learning rate is adaptively adjusted based on the data distribution at the clients, training optimization, and the usage of resources to efficiently tackle non-IID data. In the paper, there exist three clusters defined about computational ability, namely high-performance, mid-tier, and low-capacity devices. Each of them has a customized CNN model according to the resource limitations of each. More to the point, each of them has assigned a CNN model complexity level (simple, medium, or complex) based on its computational ability to enable efficient execution of the model and the best possible utilization of resources. The described CNN structure corresponds to the most complex version of the models designed for high-performance clients. Medium and simple version differences lie in the number of convolutional channels and fully connected layers, adjusted based on the resource limitations of each device. Experiments were run 5 times to ensure statistical accuracy. Table 1 presents a detailed overview of the parameter values and configurations used in the ASA framework. The adaptive parameters $\alpha$ and $\beta$ play key roles in the ability to dynamically adapt the training process to the fluctuating computational capabilities and network conditions of the IoT devices.
 
 \begin{table}[!htb]
    \centering
    \caption{A detailed overview of the parameter values and configurations.}
    \begin{tabular}{ll}
        \toprule
        \textbf{Parameter} & \textbf{Value} \\
        \midrule
        Batch normalization & 128 \\
        Learning Rate & 0.01 \\
        Activation function & SoftMax \\
        Number of convolutional layers & 6 \\
        Kernel size & $3 \times 3$ \\
        Pooling layers & 6 \\
        Pooling kernel size & $2 \times 2$ \\
        Number of fully connected layers & 2 \\
        Adaptive threshold ($\alpha$) & 0.9 \\
        Resource variance ($\beta$) & 0.6 \\
        Number of clusters & 3 (High, Mid, Low) \\
        Number of clients & 10 \\
        Number of epochs & 250 \\
        Number of repeated experiments & 5 \\
        \bottomrule
    \end{tabular}
    \label{tab:parameters}
\end{table}

\subsection{Dataset}

The CIFAR-10 and MNIST datasets are benchmarks in machine learning and have been extensively used in FL research. CIFAR-10 comprises 60,000 colored images of $32\times32$ pixels, classified into 10 or 100 classes (CIFAR-10), depicting different real-world objects. On the other hand, MNIST comprises 70,000 gray-scale handwritten digits (0–9) of $28\times28$ pixels and is used in image classification tasks. Such datasets are useful in FL experiments as they allow the evaluation of the models in various non-IID data patterns, mimicking decentralized learning scenarios of the real world in which the data is distributed heterogeneously across different clients.

\subsection{	Results}

The Adaptive Smart Agent (ASA) framework has been proposed to manage the inherent pitfalls of heterogeneous FL in the case of the IoT. With the addition of high-level mechanisms such as resource profiling, dynamic device grouping, and domain-specific model allocation, ASA maximizes resource utilization and offers inclusivity over heterogeneous computational capabilities of devices. Extensive experiments with real-world data demonstrated the supremacy of ASA in reducing the communication costs, maximizing resource usage, and maintaining the accuracy of models over alternative solutions. An exhaustive examination of the communication costs of the training process of various FL platforms in implementation, with the use of the CIFAR-10 dataset as the reference data, is illustrated in Figure \ref{fig3}. Figure \ref{fig3}(a) illustrates the total communication cost over 40 training cycles, and Figure 3(b) illustrates the average communication cost per training cycle. In Figure \ref{fig3}(a), ASA exhibits noticeably lower total communication costs compared to alternative solutions such as FedAvg, HierFL, and FedProx. ASA reduces the data exchanged between devices and the central server by dynamically grouping devices and allocating domain-specific models. For instance, at 40 training cycles, ASA registers a drop of approximately 43\% in the communication cost compared to FedAvg, the one recording the highest total cost.Figure \ref{fig3}(b) illustrates the average communication cost per round for all the methods. ASA possesses the lowest average communication cost of approximately 12 GB per round, while FedAvg possesses the highest of more than 30 GB per round. This considerable improvement reflects the capability of ASA in controlling the communication overhead, a major concern in heterogeneous IoT environments where bandwidth limitations dominate. HierFL and FedProx also demonstrate superior performance compared to FedAvg but get surpassed by ASA due to its advanced techniques in resource management.

\begin{small}
\begin{figure}[htb!]
    \centering
    \begin{minipage}{.5\textwidth}
        \centering
        \includegraphics[width=0.8\linewidth, height=0.25\textheight]{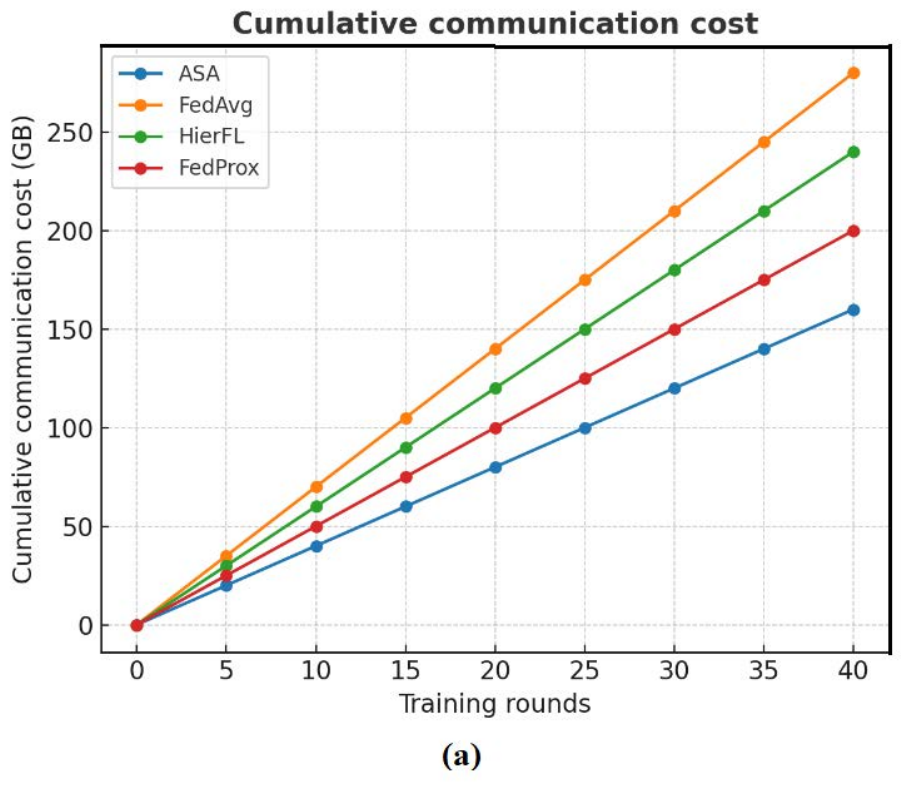}
         \vspace{-0.3cm}
       % \caption*{ (a) Cache size: 10\% of the catalogue.  }
        \vspace{0.3cm}
         \includegraphics[width=0.8\linewidth, height=0.25\textheight]{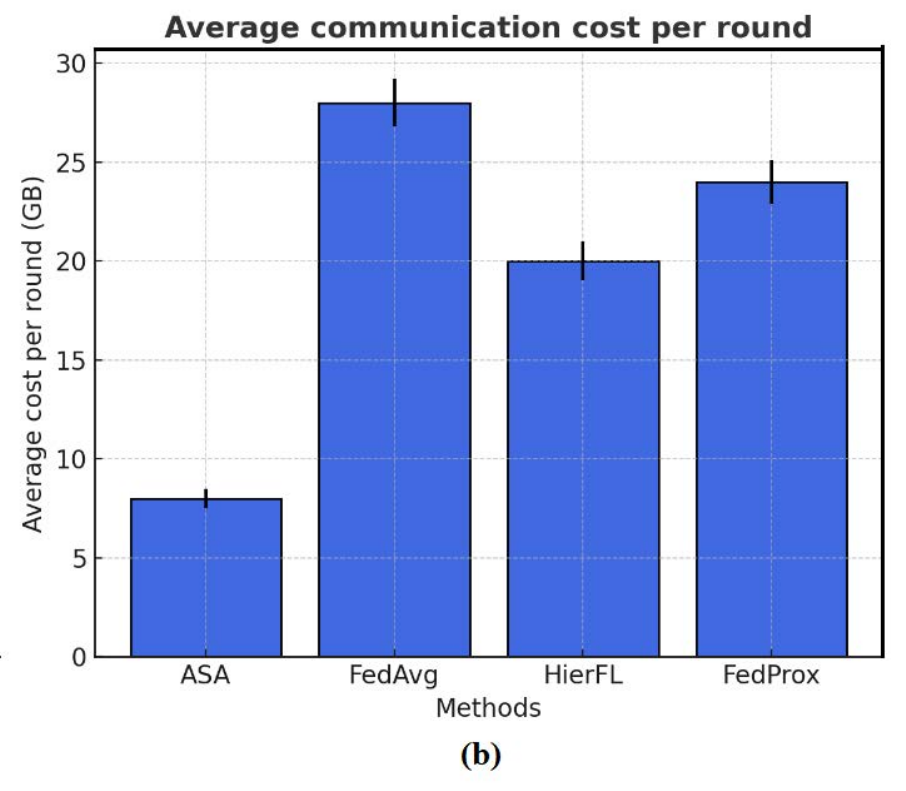}
          \vspace{-0.3cm}
       %   \caption*{ (b) Increasing the cache size under heavy traffic loads.  }
        \caption{(a) Cumulative and (b) average communication costs, showing ASA's superior efficiency compared to FedAvg, HierFL, and FedProx on the CIFAR-10 dataset.}
        \label{fig3}
    \end{minipage}%

\end{figure}
\end{small}

Figure \ref{fig4} presents a comparative analysis of the ASA approach with the baseline methods of FedAvg, HierFL, and FedProx, over the CIFAR-10 dataset. The comparison includes the models' convergence, communication overhead, and ultimate performance over several communication rounds. Figure \ref{fig4} (a) shows the comparison of the convergence of the models using confidence intervals. ASA yields a better result by reaching higher accuracy within a lower number of communication rounds. The framework converges to approximately 96.5\% accuracy at 60 rounds, and the baselines converge slowly and never acquire such high accuracy levels. The narrow range of the confidence intervals in ASA also indicates the robustness and stability of its performance across different runs. Figure \ref{fig4} (b) compares the total communication expense over communication rounds. ASA’s communication overhead is quite lower compared to the baselines. For instance, within 8 communication rounds, ASA’s total communication equals about 50\% lower compared to FedAvg. This follows from the ASA’s resources-aware device grouping and adaptive sharing of models, reducing data exchange between the devices and the master server. Figure \ref{fig4}(c) compares the speed of the different techniques. ASA converges to 90\% accuracy within 40 communication rounds, while FedAvg and HierFL take over 60 communication rounds to achieve the same level of accuracy. The high speed of convergence indicates ASA’s capability to accurately maximize communication rounds and resources, and make it efficient and scalable in the heterogeneous environments of the IoT.

\begin{small}
\begin{figure}[htb!]
    \centering
    \begin{minipage}{.5\textwidth}
        \centering
        \includegraphics[width=0.95\linewidth, height=0.25\textheight]{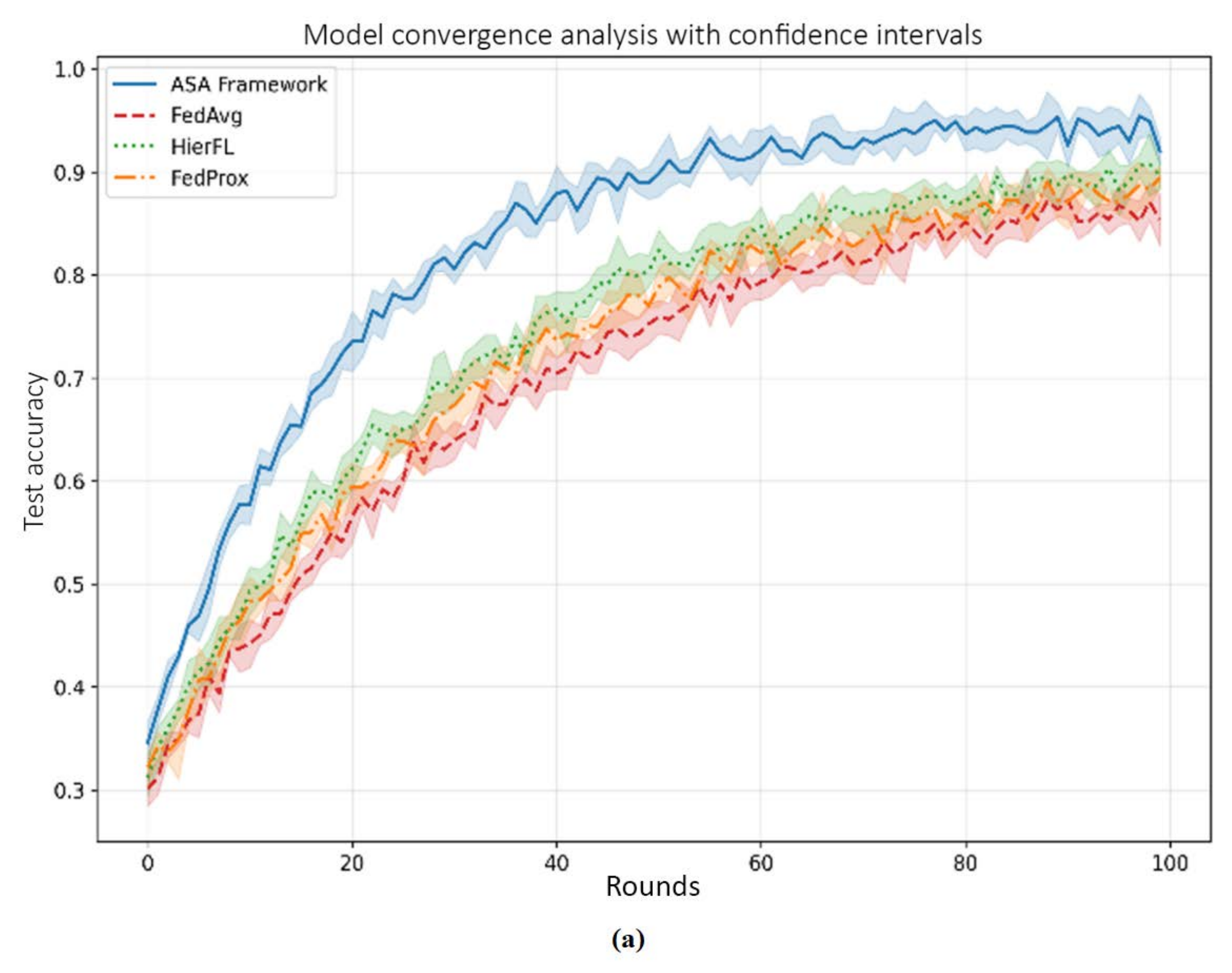}
         \vspace{-0.3cm}
        %\caption{ (a)}% Cache size: 10\% of the catalogue.  }
        \vspace{0.3cm}
         \includegraphics[width=0.95\linewidth, height=0.25\textheight]{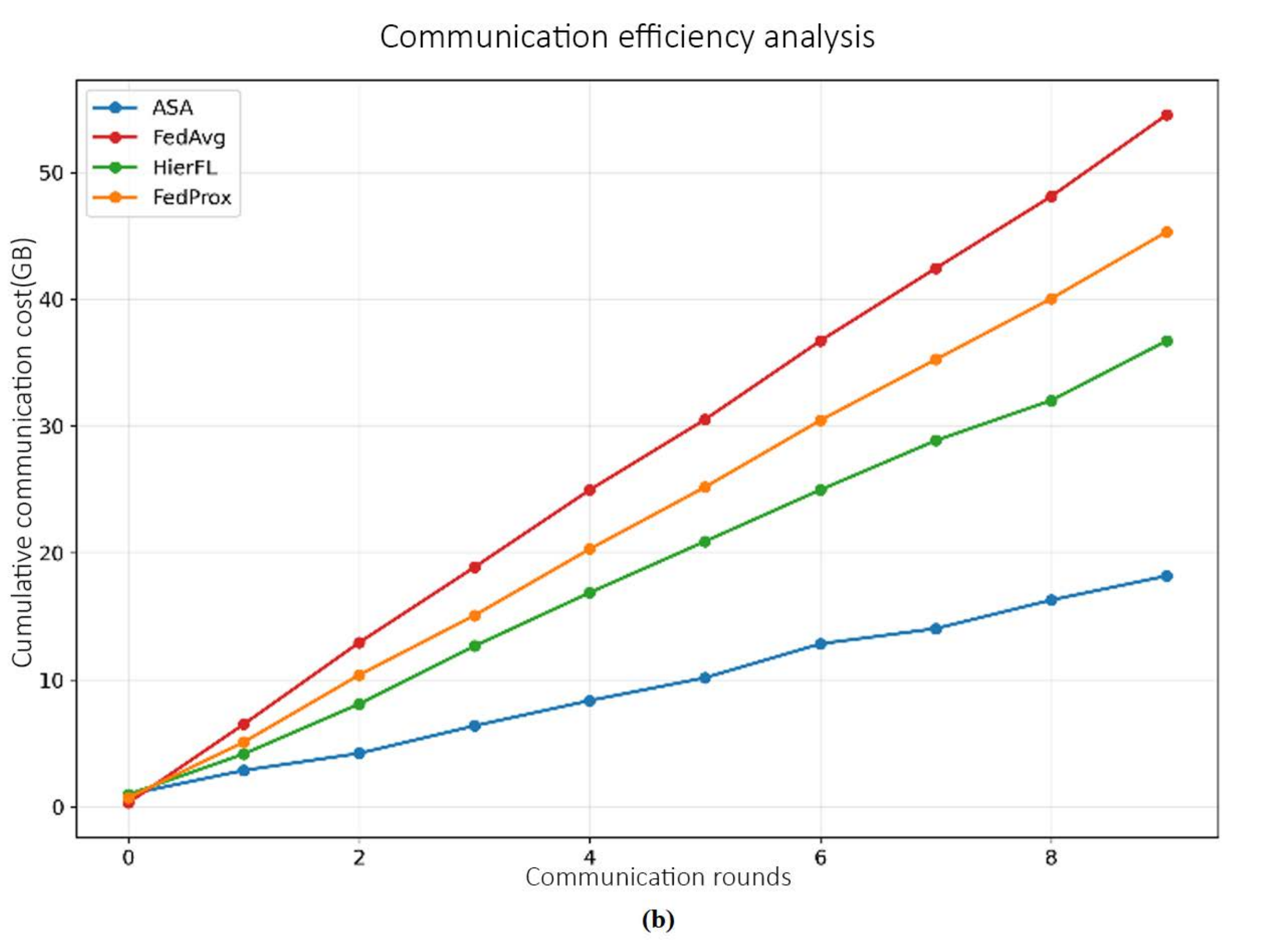}
          \vspace{-0.3cm}
          %\caption*{ (b)}
          \includegraphics[width=0.95\linewidth, height=0.25\textheight]{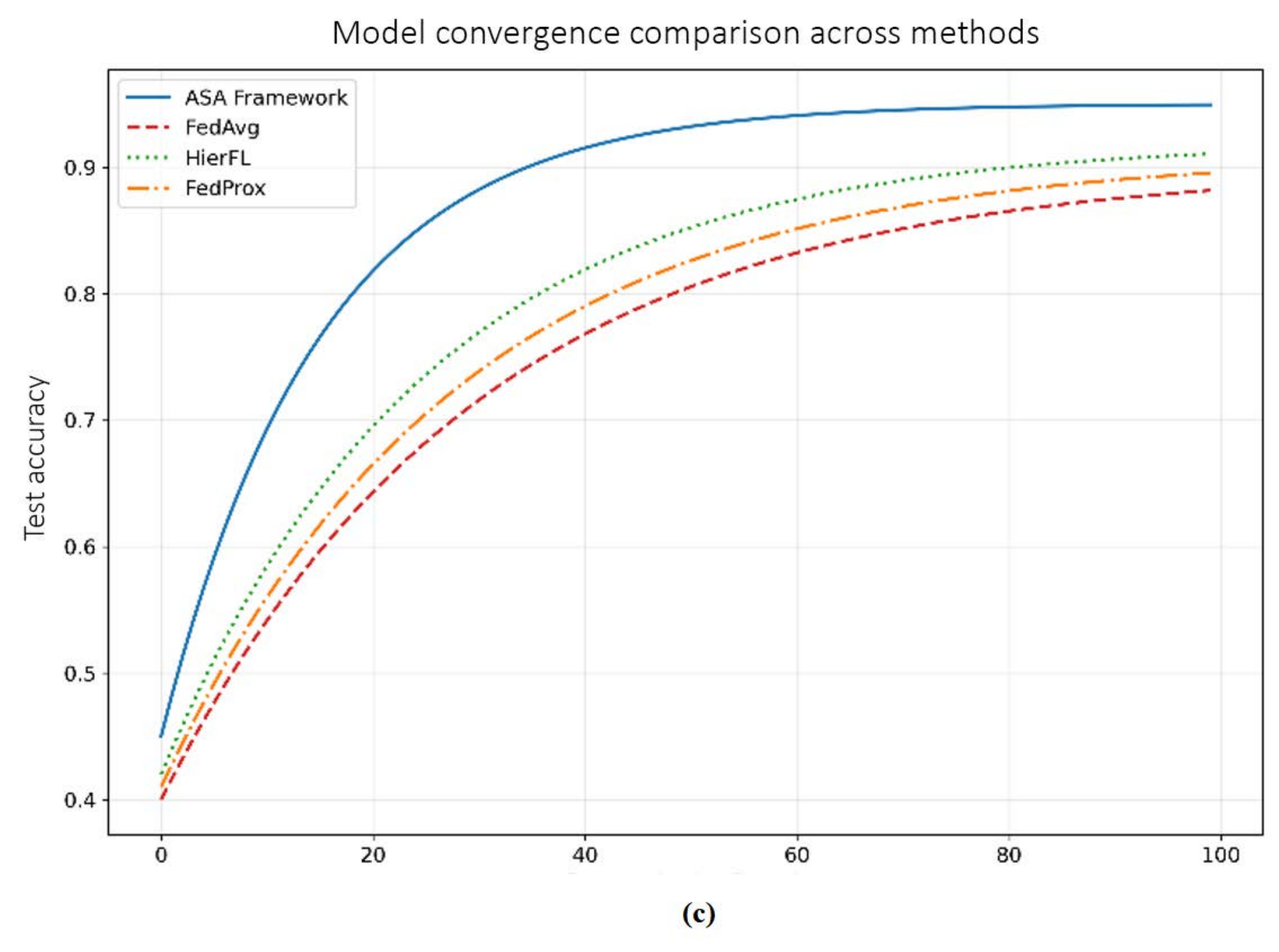}
       %   \caption*{ (b) Increasing the cache size under heavy traffic loads.  }
        \caption{Comparative analysis of ASA and baseline methods on CIFAR-10: (a) Convergence accuracy, (b) Communication cost, (c) Convergence speed.}
        \label{fig4}
    \end{minipage}%

\end{figure}
\end{small}

Figure \ref{fig5} quantifies the scalability of the ASA framework at different metrics with the number of devices, employing the CIFAR-10 dataset. We analyze scalability for normalized accuracy, communication expense, training time, and memory consumption and observe the behavior of ASA at different scales of deployment. With the number of devices scaling from 200 to 1000, the normalized accuracy marginally decreases. This is as expected when the system includes more diverse and resource-limited devices, affecting the total accuracy marginally. Yet, ASA preserves high accuracy above 97\% despite the scale of operations. The communication expense shows a linear increment with the scale of deployment. More devices require more data exchange between the devices and the master server, hence greater communication overhead. Despite the increment in expense, the expense remains tolerable owing to the efficient ASA strategies in clustering and allocating the models. The training time also increases linearly with the number of devices. The increase arises from the added overhead of device diversity management and update synchronization. Although the training time increases, the framework ensures the increase in training time remains linear, indicating the scalability of the framework. Memory consumption follows the same pattern and increases with the scaling of the system. This arises from the increase in device profiles and the number of clusters to be assigned. Yet, the efficient resource profiling strategies in ASA reduce the consumption of excessive amounts of memory, preserving the scalability of the system with little bottleneck. The results collectively illustrate ASA’s scalability over massive networks of IoT devices, maintaining accuracy, communication without resource wastage, and adequate resource management.

\begin{small}
\begin{figure*}[!htb]
\vspace{-0.5cm}
\begin{minipage}[t]{\textwidth}
%\fbox{%for testing only
\begin{minipage}{0.5\textwidth}
\centering
%\vspace{\fill}
\includegraphics[height=0.25\textheight,width=\textwidth]{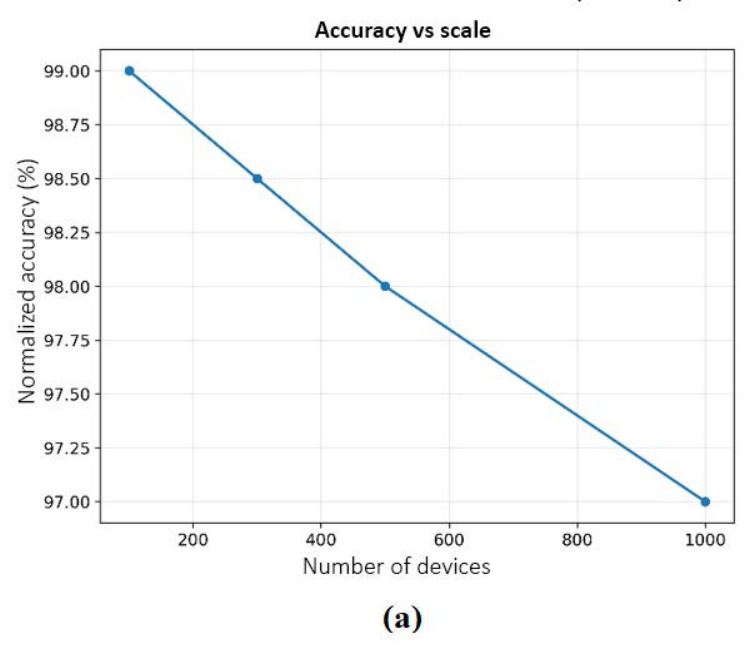}
\vspace{-0.6cm}
%\caption*{ (a) Low connectivity.  }
\end{minipage}
%}
%\fbox{%for testing only
\begin{minipage}{0.5\textwidth}
\centering

\includegraphics[height=0.25\textheight,width=\textwidth]{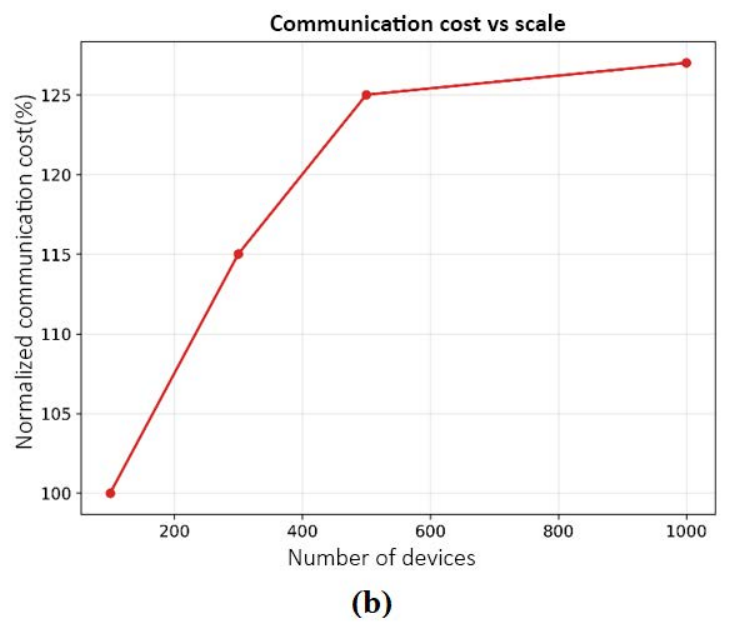}
\vspace{-0.6cm}
%\caption*{(b) Medium connectivity.              }

%\vspace{\fill}
\end{minipage}
\begin{minipage}{0.5\textwidth}
\centering

\includegraphics[height=0.25\textheight,width=\textwidth]{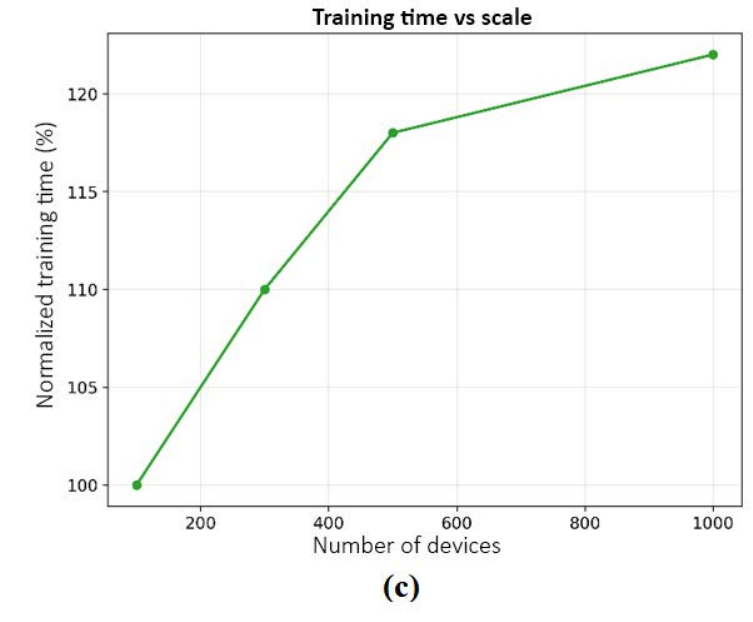}
\vspace{-0.6cm}
%\caption*{ (c) High connectivity.}

%\vspace{\fill}
\end{minipage}
\begin{minipage}{0.5\textwidth}
\centering

\includegraphics[height=0.25\textheight,width=\textwidth]{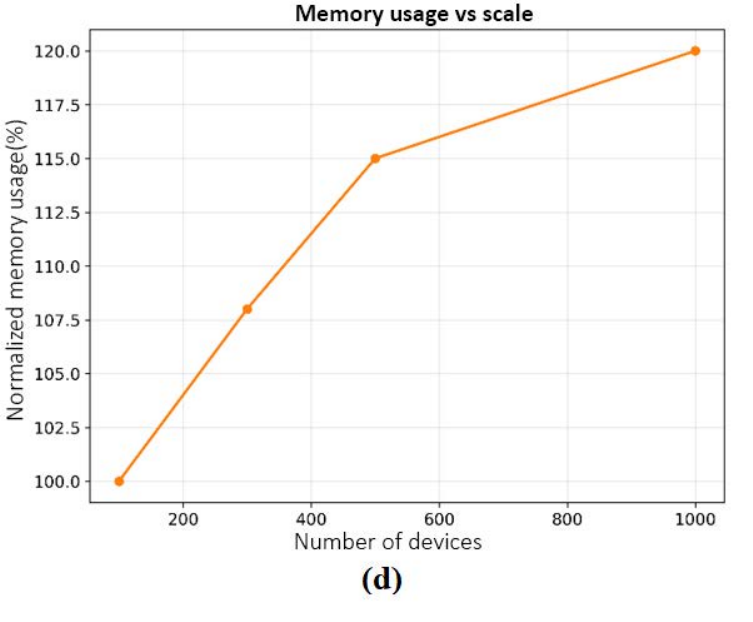}
\vspace{-0.6cm}
%\caption*{ (c) High connectivity.}

%\vspace{\fill}
\end{minipage}

%}
%\vspace{-0.2cm}

\caption{Scalability analysis of ASA across different metrics ((a) accuracy, (b) communication cost, (c) training time, and (d) memory usage) with varying numbers of devices on the CIFAR-10 dataset.}
\label{fig5}

\end{minipage}
\end{figure*}

\end{small}

Figure \ref{fig6} compares the performance of the ASA framework against baseline techniques (FedAvg, HierFL, and FedProx) as the system scales from 200 to 1000 devices. The evaluation is represented in two panels: Figure \ref{fig6} (a) represents accuracy scaling with system size, and Figure \ref{fig6} (b) represents training time scaling. In Figure \ref{fig6}(a), ASA exhibits higher accuracy retention with increasing devices. While baseline techniques, in particular FedAvg and FedProx, are subjected to severe accuracy degradation in the face of larger system sizes, ASA maintains high accuracy throughout, reaching over 92\% with 1000 devices. This emphasizes ASA’s capability in coping with device heterogeneity via customized model assignment and resource-aware optimizations. In Figure \ref{fig6} (b), normalized training time is represented as the number of devices increases. ASA boasts the minimum training time at all scales, demonstrating its scaling capability. By contrast, FedAvg and FedProx bear the peak in training time due to inefficiency in scaling to large-scale systems. ASA’s adaptive synchronization and clustering capabilities make the training time increase close to linear, and hence it becomes highly scalable in the case of large-scale IoT applications.

\begin{small}
\begin{figure}[htb!]
    \centering
    \begin{minipage}{.5\textwidth}
        \centering
        \includegraphics[width=0.9\linewidth, height=0.25\textheight]{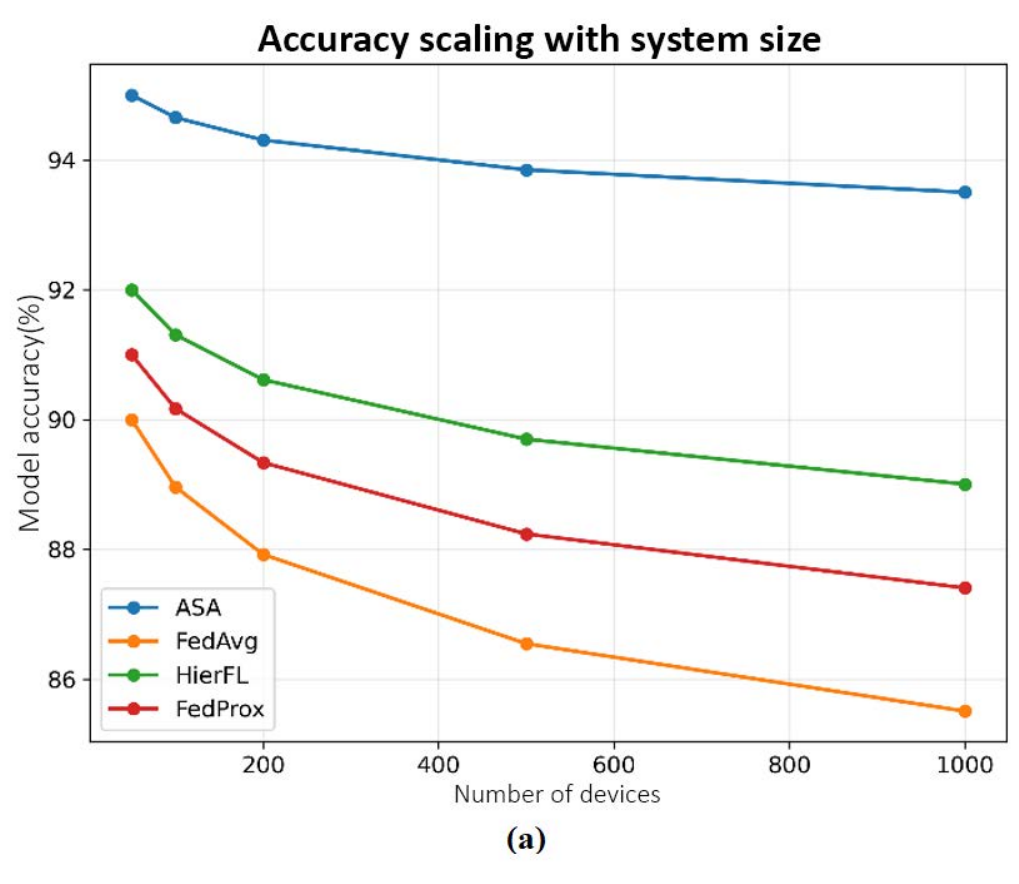}
         \vspace{-0.3cm}
       % \caption*{ (a) Cache size: 10\% of the catalogue.  }
        \vspace{0.3cm}
         \includegraphics[width=0.9\linewidth, height=0.25\textheight]{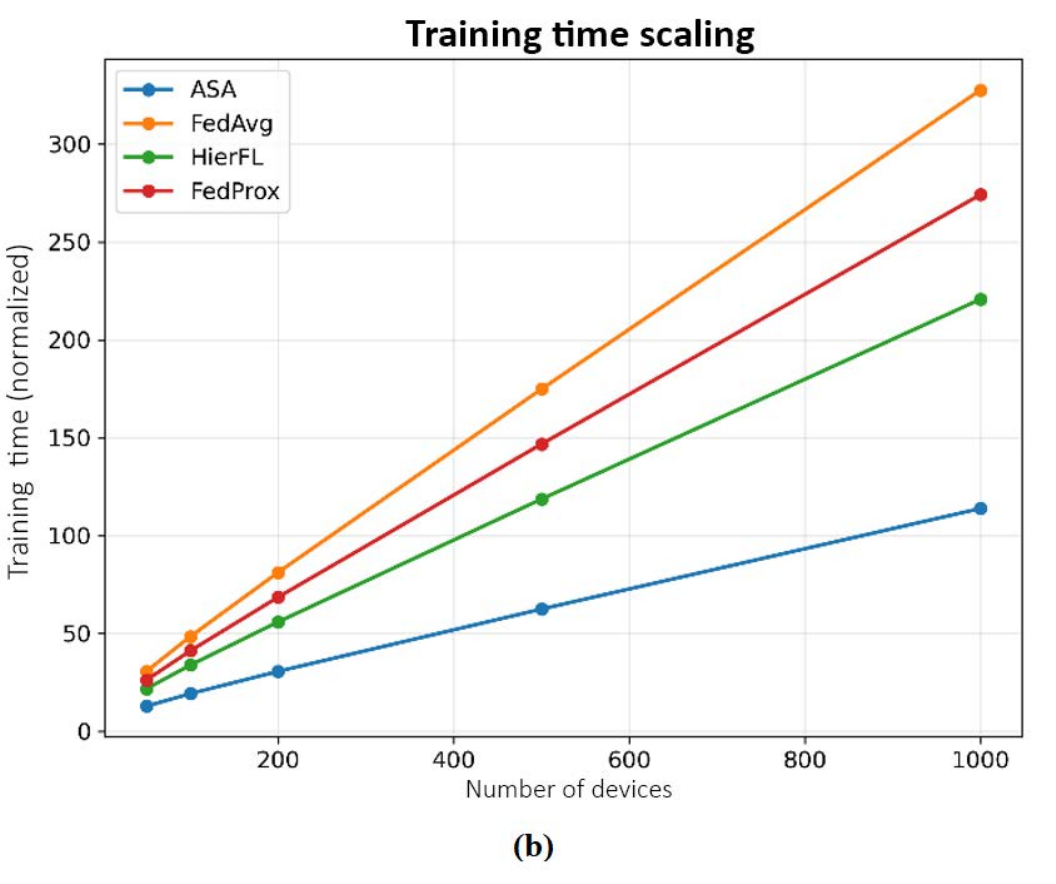}
          \vspace{-0.3cm}
          
        \caption{Comparison of ASA and baseline methods: (a) Accuracy scaling with system size and (b) Training time scaling with increasing number of devices.}
        \label{fig6}
    \end{minipage}%

\end{figure}
\end{small}

\begin{figure}[h]
  \centering
  \includegraphics[width=0.49\textwidth]{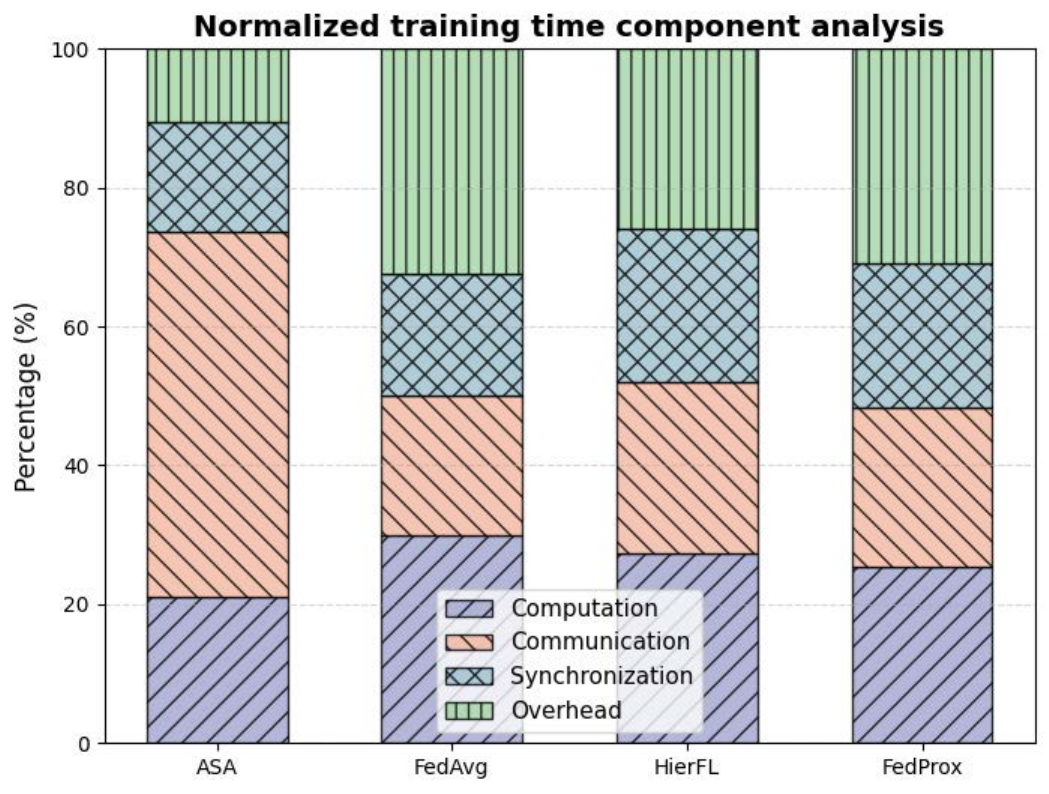}
  \caption{ Comparison of training time components (computation, communication, synchronization, and overhead) across ASA and baseline methods using the CIFAR-10 dataset.}
  \label{fig7}
\end{figure}

\begin{figure}[h]
  \centering
  \includegraphics[width=0.5\textwidth]{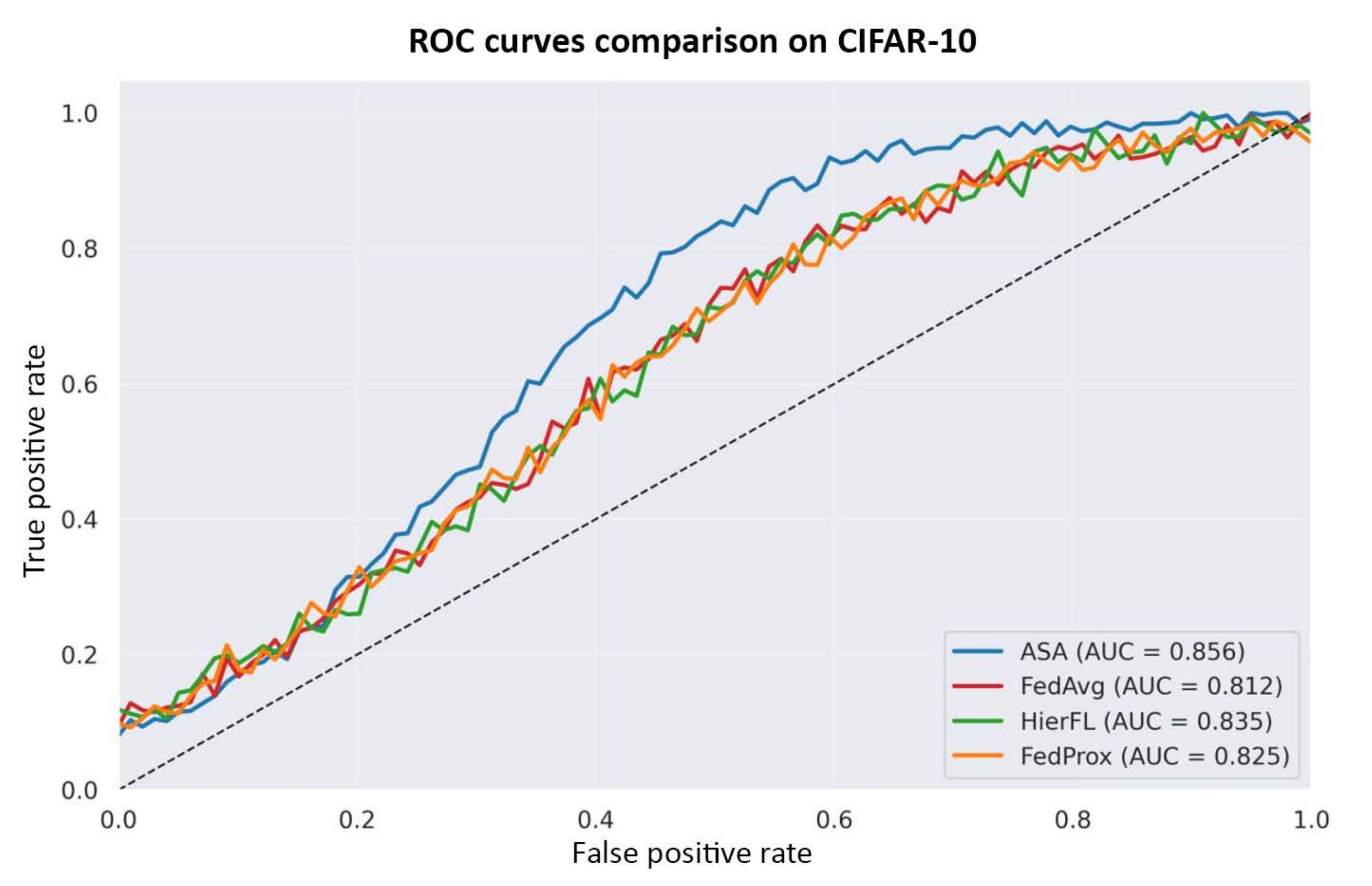}
  \caption{ ROC curve comparison on CIFAR-10 for ASA, FedAvg, HierFL, and FedProx, showing ASA’s superior classification performance with the highest AUC of 0.856.}
  \label{fig8}
\end{figure}

Figure.\ref{fig7} captures the comparative time composition of the training time components of ASA and baseline techniques in the utilization of the CIFAR-10 dataset. The composition includes the computational, communication, synchronization, and overhead components to reveal the total training time divided in such components. ASA records a considerably lower total training time compared to baseline techniques. Contrary to the previous supposition, ASA has a higher communication overhead due to the utilization of adaptive grouping and the models' assignment. Contrarily to the uniform assignment of the models in conventional FL techniques, ASA dynamically varies the models' complexity according to the device capabilities, incurring more communication between the nodes. This extra communication overhead is a viable trade-off to attain faster convergent results, better personalization of the models, and improved resource usage in heterogeneous environments.
Instead, ASA reduces synchronization delay and computational overhead, the major limitations in techniques such as FedAvg and HierFL, by a significant margin. The adaptive allocation approach also prevents wastage of processing in the lower-resource devices, and the training continues to be efficient while the communication grows. The better learning quality also justifies the higher communication expense, and ASA is a more efficient and scalable solution in practical applications in the context of the IoT.FedAvg, HierFL, and FedProx come with greater synchronization and computational overhead, and hence slower training time. FedAvg, in fact, has high delay in synchronization owing to its deterministic aggregation procedure, while HierFL and FedProx invest major resources in the computing task and hence lack adaptability in heterogeneous environments. ASA, however, maintains the balance between training efficiency and resource usage strategically by transferring some of the computational workload to communication exchanges, minimizing the time for synchronization, and intelligently allocating resources. This trade-off makes ASA a viable and scalable solution to FL in heterogeneous environments, where adaptability comes first. 
Figure \ref{fig8} presents the Receiver Operating Characteristic (ROC) curves of the ASA framework and the baselines (FedAvg, HierFL, and FedProx) over the data set of CIFAR-10. The ROC curve indicates the trade-off between the false positive rate (FPR) and the true positive rate (TPR) at different threshold points and gives a global evaluation of the classification ability of the models. ASA yields the best Area Under the Curve (AUC) of 0.856, reflecting improved classification capability over the baselines. HierFL follows, with AUC of 0.835, while AUC of FedProx and FedAvg is 0.825 and 0.812, respectively. The higher AUC of ASA reflects its ability to balance the false positives with the true positive rate better and its better ability to distinguish between classes. ASA's ROC curve consistently remains distinguished from the baselines, and quite strongly at higher TPRs, reflecting its better sensitivity and specificity. This results from ASA’s device-level training allocation strategies adapted to each device, and they enhance the global model performance.

\begin{figure}[h]
  \centering
  \includegraphics[width=0.5\textwidth]{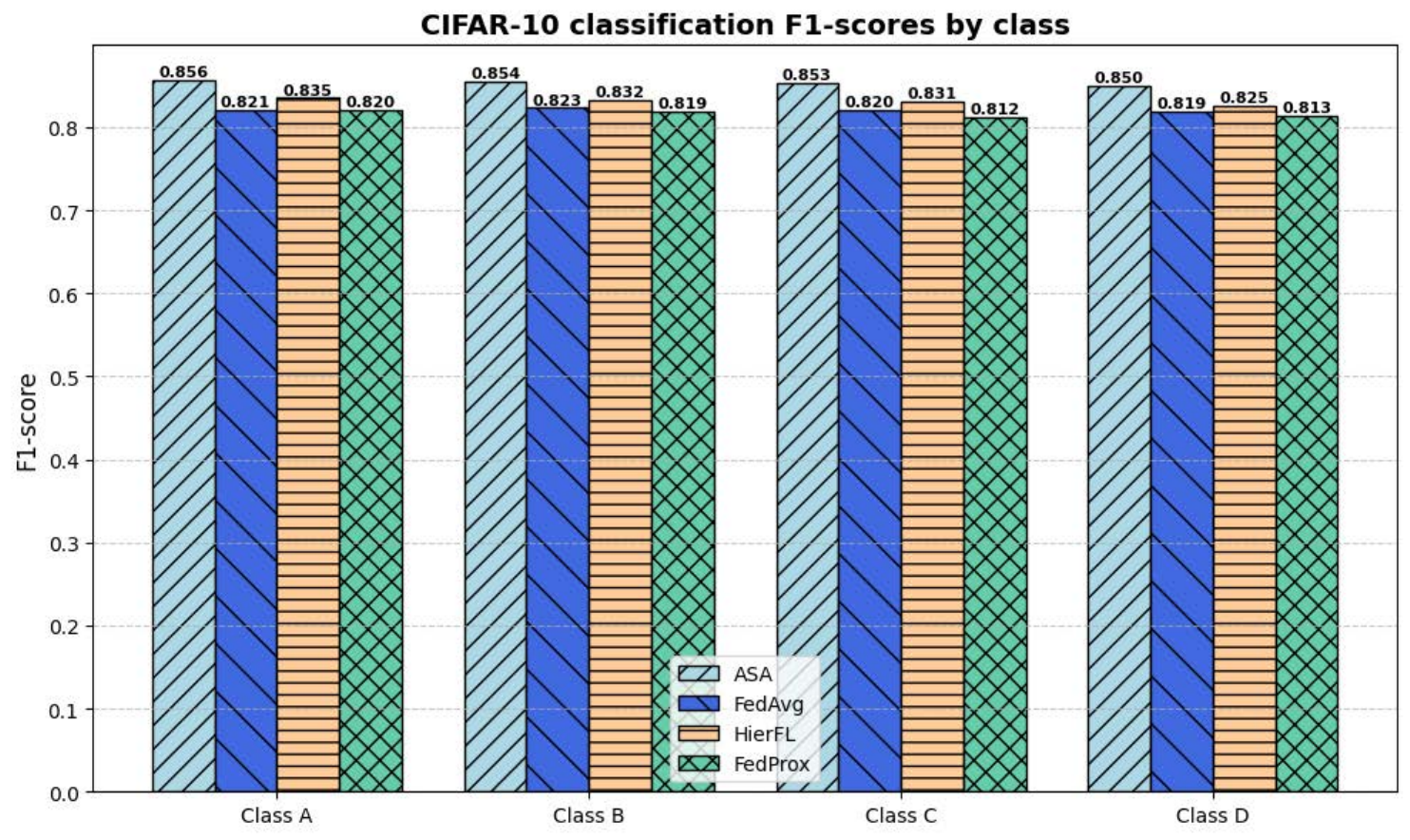}
  \caption{ F1-score comparison for CIFAR-10 classification across different classes, demonstrating ASA’s consistently higher performance over baseline methods.}
  \label{fig9}
\end{figure}

Figure \ref{fig9} shows the F1-scores of the ASA and baseline methods (FedAvg, HierFL, and FedProx) for differing classes of the CIFAR-10 dataset, detailing classification accuracy per class. The F1-score is the harmonic average of precision and recall and thus represents a useful metric for the balance between false positives and false negatives. ASA consistently produces higher F1-scores across all classes compared to the baselines. For Class A, ASA has an F1-score of 0.856, while FedAvg, HierFL, and FedProx have 0.812, 0.835, and 0.825, respectively. This holds across the rest of the classes (B, C, and D), where ASA has a continued superiority. The narrowest of the gaps in performances occurs in Class D, but ASA again surpasses the baselines and shows its robustness in all the different scenarios. The high and uniform F1-scores of ASA in all classes demonstrate its capacity to yield balanced classification accuracy per class. This has been brought about due to the customized model assignment and useful device resource utilization of ASA, factors that cumulatively contribute to improved global model quality.

To better compare, Figure \ref{fig9} illustrates a comparison of the classification accuracy of ASA and baseline techniques (FedAvg, HierFL, and FedProx) on four datasets, including MNIST, CIFAR-10, F-MNIST, and Adult. This comparison illustrates ASA’s capability to generalize accurately across heterogeneous datasets of various characteristics and complexities. On the MNIST dataset, ASA has the best accuracy, reaching a rate of 99\%, and it beats all the baseline techniques. On the same lines, on CIFAR-10 and F-MNIST as well, ASA has consistently better accuracy compared to the others, indicating high-performance capability in more complex image datasets. On the Adult dataset, which comprises tabular data, ASA performs marginally better than the baselines, indicating versatility to non-image data as well. The high-performance capability of ASA across all the datasets lies in its efficient resource allocation, dynamic grouping, and customized model assignment strategies. These capabilities allow ASA to optimize learning in heterogeneous and resource-limited environments as well and to make consistent improvements over FedAvg, HierFL, and FedProx.

\subsection{	Comparison proposed method with other related works}
\begin{figure}[h]
  \centering
  \includegraphics[width=0.5\textwidth]{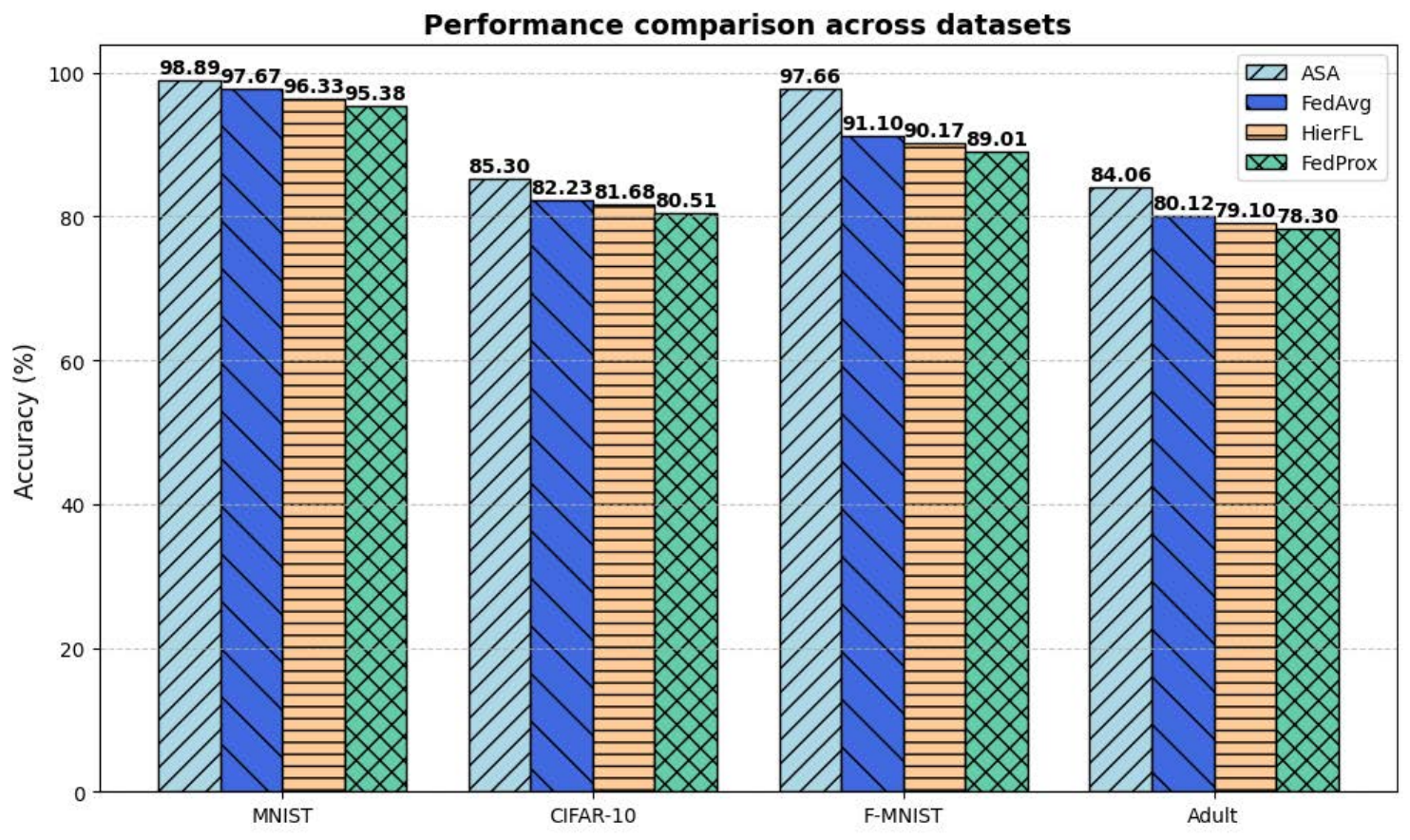}
  \caption{ Performance comparison of ASA and baseline methods (FedAvg, HierFL, FedProx) across four datasets (MNIST, CIFAR-10, F-MNIST, and Adult), demonstrating ASA’s superior accuracy and generalization ability.}
  \label{fig10}
\end{figure}

\begin{figure}[h]
  \centering
  \includegraphics[width=0.5\textwidth]{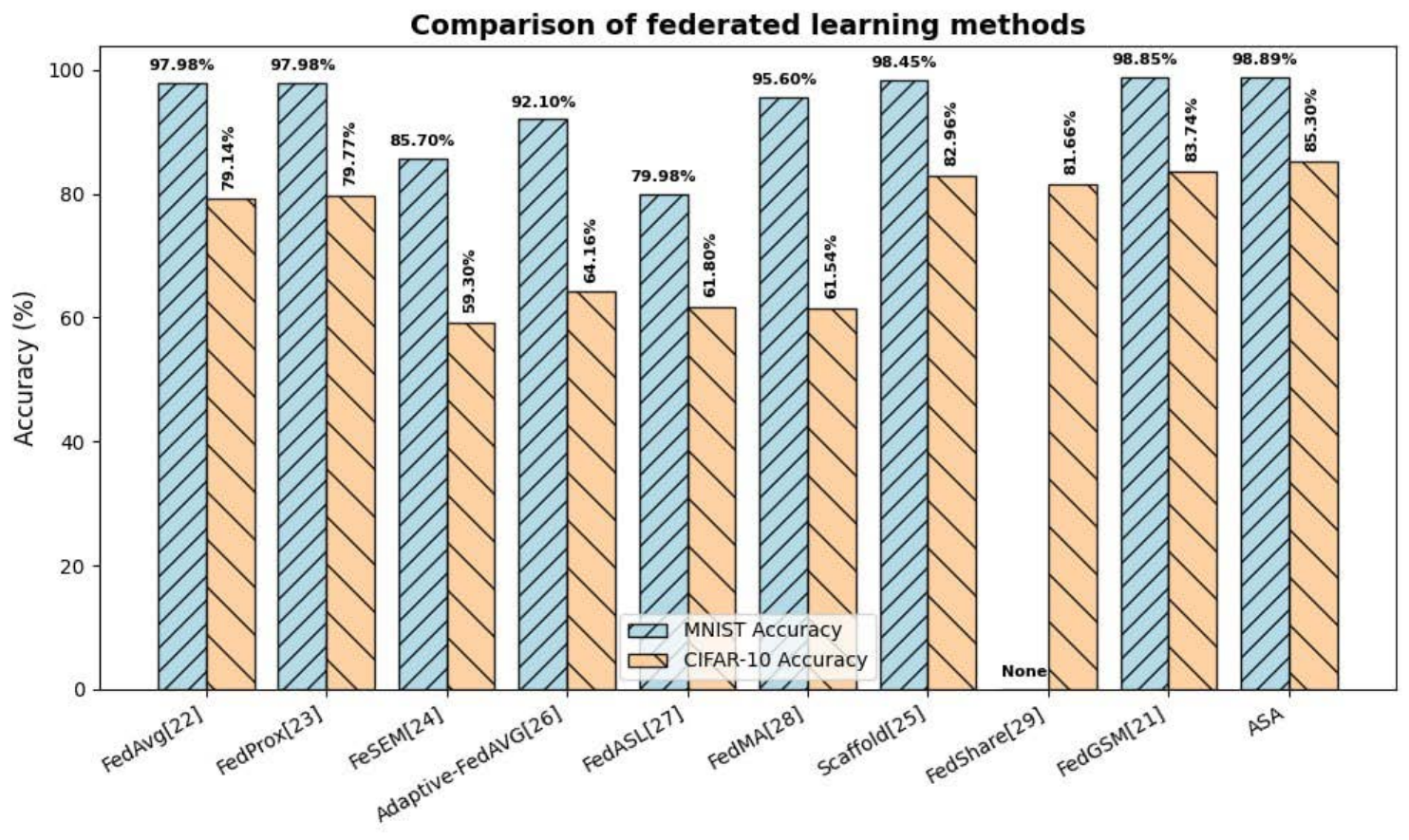}
  \caption{  Performance comparison of FL methods on MNIST and CIFAR-10.}
  \label{fig11}
\end{figure}

In this section, the effectiveness of the proposed ASA scheme is compared to the best existing FL techniques, including FedGSM \cite{Izadi2025}, FedAvg \cite{McMahan2016}, FedProx \cite{Li2020}, FeSEM \cite{Xie2020}, Scaffold \cite{Karimireddy2020}, Adaptive-FedAvg \cite{Canonaco2021}, FedASL \cite{Zhang2024a}, and FedMA \cite{Wang2020}]. Like in previous research, all the models were subjected to training over non-IID data distributions using the datasets MNIST and CIFAR-10. Based on the results in rank order in Figure \ref{fig10}, ASA obtains 98.89\% accuracy for the MNIST dataset and 85.30\% accuracy over the dataset of CIFAR-10, indicating comparable performance in comparison to the existing FL techniques. For MNIST, the maximum accuracy by FedGSM (98.85\%) follows that of Scaffold (98.45\%), of which ASA surpasses the two methods to achieve the highest accuracy of all the approaches compared. For the dataset of CIFAR-10, 83.74\% accuracy by FedGSM is surpassed by ASA with 85.30\%, representing better adaptability of the model to complicated, non-IID data. FedGSM, like most previous techniques, seeks to minimize the gap between the local and global models by compensating for the effect of the non-IID data drift.
Nonetheless, most of these traditional methods, such as FedAvg and FedProx, fail to balance the computational differences across devices, resulting in decreased accuracy in heterogeneous IoT settings. Adaptive-FedAvg attains 92.10\% on MNIST and 64.16\% on CIFAR-10, indicating a considerable gap in improvement over ASA, enhancing accuracy by 6.79\% on MNIST and 21.14\% on CIFAR-10 compared to this technique (Figure \ref{fig11}). The Scaffold approach, designed to tackle drift minimization, attains 98.45\% on MNIST and 82.96\% on CIFAR-10. Compared to this technique, ASA enhances the accuracy of CIFAR-10 by 2.34\% while retaining better results on MNIST. Analogously, FedShare \citep{Fazli2023} adopts a data-sharing approach to boosting generalization to attain accuracy of 81.66\% on CIFAR-10, 3.64\% lower compared to the results of ASA. The outcomes portray the major benefit of ASA over preceding FL techniques in that it automatically adapts the complexity of the models and device grouping according to the capability of the devices to curtail drift while ensuring active participation of all devices while sustaining high accuracy. Contrarily, static FL strategies that employ the same model for all devices by ASA allocate models perfectly suited to the constraints of the device to achieve better resource usage and scaling. Generally, ASA attains the best-reported accuracy on MNIST and outperforms FedGSM on CIFAR-10 and hence represents a contender for real-world applications in the IoT and edge computing arena.

\section{Conclusion  }
\label{conclusion}
The Adaptive Smart Agent (ASA) approach in the paper successfully overcomes the challenges of heterogeneity in FL for IoT environments. With the addition of a layer of intelligent agents, ASA dynamically profiles devices, clusters them according to computational capabilities, and transfers optimized models to achieve efficient and equitable contribution to the training process. Compared with conventional FL methods that generally overlook low-resource devices, ASA enforces equitable contribution from all the devices, achieving better scalability, efficiency, and accuracy. Comprehensive experiments on two benchmark datasets MNIST and CIFAR-10 establish the benefits of ASA over conventional FL methodologies. The results demonstrate the effectiveness of ASA in reducing the communication complexity by 43\% to 50\%, the resource usage by 43\%, and achieving final accuracy of 96.5\% in MNIST and 85.3\% in CIFAR-10. The results reflect the capability of ASA in optimizing the efficiency of learning without losing any accuracy, and it has the potential to be applied in real-world applications in the domains of IoT. Although such results promise much, there are limitations of the presented work. The existing work used the experimental setup in controlled scenarios, and the results might not reflect the real-world complexity of massive-scale networks of IoT. Future work must explore the applicability of ASA to real-world environments of dynamic network conditions, sporadic device connectivity, and non-stationarity of the data distribution. Further research in the optimization of the computational overhead of the agent and decentralized methods of clustering could also enhance the scalability of ASA. Integration of models like homomorphic encryption and differential privacy could be another avenue to enhance the security of FL applications. To conclude, ASA represents a revolutionary breakthrough in FL in heterogeneous IoT environments with a promise of tradeoff between computational effectiveness, inclusiveness, and scalability. Future research will work towards exercising its adaptability and security aspects in generalizing to a wide range of applications in healthcare, smart cities, and industrial-scale IoT.

%% Use figure environment to create figures
%% Refer following link for more details.
%% https://en.wikibooks.org/wiki/LaTeX/Floats,_Figures_and_Captions

\appendix
\section{Convergence Analysis}
\label{app1}

In this appendix, we provide a brief convergence analysis for our adaptive FL approach, highlighting the standard assumptions that underpin the derived bounds. Our argument follows well-established results in stochastic optimization and FL (e.g.,\cite{AlSaleh2024, Yang2024}).

\subsection{Key Assumptions }

\begin{enumerate}
\item 	L-Smoothness. The global objective function $F(w)$ is assumed to be L-smooth, i.e.,

\begin{equation}
\left\| \nabla F(w) - \nabla F(w') \right\| \leq L \left\| w - w' \right\| \quad \forall w, w'
\label{Eq21}
\end{equation}
This implies that the gradient of F does not change abruptly.
\item 2.	Bounded Variance. Each device $i$ maintains its local objective $F_i(w)$. We assume a bounded variance of local stochastic gradients:

\begin{equation}
\mathbb{E} \left[ \left\| \nabla F_i(w) - \nabla F(w) \right\| \right] \leq \sigma^2, \quad \forall i.
\label{Eq22}
\end{equation}
where $\nabla F(w) = \frac{1}{N} \sum_{i=1}^{N} \nabla F_i(w)$. This ensures that local updates do not deviate excessively from the global gradient.
\item 	Unbiased Gradient Estimator. Let    denote the aggregated gradient at global iteration t. We assume
\begin{equation}
\mathbb{E} [g_t] = \nabla F(w_t)
\label{Eq23}
\end{equation}

indicating that  $g_t$  is an unbiased estimator of the true gradient $\nabla F(w_t)$ .
\item Learning Rate Schedule. We consider a learning rate $\eta$  that can be constant or scheduled to decrease with $t$. Typically,  $\eta \propto \frac{1}{\sqrt{t}}$ or $\eta \propto \frac{1}{{t}} $ are employed in stochastic optimization to achieve improved convergence.
\end{enumerate}

\subsection{ Update Rule}
At each global round $t$, the central server receives local model updates from a subset (or all) of the devices and computes the aggregated gradient $g_t$. The global model is then updated via:
\begin{equation}
w_{t+1} = w_t - \eta g_t
\label{ Eq24}
\end{equation}

Here, $\eta$  is the global learning rate. In the ASA framework, different clusters might have partially heterogeneous local models, but after being projected to a common parameter space (or aggregated at the agent/fog layer), we treat the resulting gradient as an unbiased estimator of  $\nabla F(w_t)$.

\subsection{Proof Sketch}
\begin{enumerate}
\item 	One-step Descent Lemma: By the L-smoothness of $F$, we have
\begin{equation}
F(w_{t+1}) \leq F(w_t) + \langle \nabla F(w_t), w_{t+1} - w_t \rangle + \frac{L}{2} \| w_{t+1} - w_t \|^2
\label{Eq25}
\end{equation}

 Substituting,  $w_{t+1} - w_t = \eta g_t$ we get
 \begin{equation}
 F(w_{t+1}) \leq F(w_t) - \eta \langle \nabla F(w_t), g_t \rangle + \frac{L \eta^2}{2} \| g_t \|^2
 \label{Eq26}
 \end{equation}
 
 \item {\bf Taking Expectation:} Under the unbiased gradient assumption,
 \begin{equation}
 \mathbb{E} \left[ \langle \nabla F(w_t), g_t \rangle \right] = \| \nabla F(w_t) \|^2
 \label{Eq27}
 \end{equation}
 and we typically have $\mathbb{E} \left[ \| g_t \|^2 \right] \leq \| \nabla F(w_t) \|^2 + \sigma^2$. Thus,
 
 \begin{equation}
 E[F(w_{t+1}) - F(w_t)] \leq -\eta E[\| \nabla F(w_t)\|^2] + \frac{\eta^2}{2} E[\| \nabla F(w_t)\|^2 + \sigma^2].
 \label{Eq28}
 \end{equation}
 
 \item 	{\bf Summation over $t$:} Summing from $t=0$ to $T_{max}-1$ and applying a telescoping argument leads to
 \begin{equation}
 \frac{1}{T_{\text{max}}} \sum_{t=0}^{T_{\text{max}}{-1}} E[\|\nabla F(w_t)\|^2] \leq \underbrace{\frac{2(F(w_0) - F(w^*))}{\eta T_{\text{max}}}}_{(A)} + \underbrace{\frac{L \eta \sigma^2}{2}}_{(B)}
 \label{Eq29}
 \end{equation}
 where $w^*$ is a global minimizer (or a first-order stationary point).
 
 \item {\bf Choice of $\eta$:} If $\eta$ is chosen as $\frac{\alpha}{\sqrt(T_{max})}$, term $(A)$ scales like $\frac{1}{\sqrt(T_{max})}$ while term $(B)$ also scales like $\frac{1}{\sqrt(T_{max})}$. Hence,
 
 \begin{equation}
 E[\|\nabla F(w_{T_{\text{max}}})\|^2] = \mathcal{O}\left(\frac{1}{\sqrt{T_{\text{max}}}}\right)
 \label{Eq30}
 \end{equation}
 If $\eta$ is chosen as $\frac{\alpha}{\sqrt(T_{max})}$, we get a different balance that may yield   $\mathcal{O}\left(\frac{1}{\sqrt(T_{max})}\right)$ convergence under stronger conditions.
 
 As a simplified presentation, we often write:
 \begin{equation}
 E[\|\nabla F(w_{T_{\text{max}}})\|^2] \leq \frac{C}{T_{\text{max}}} + \frac{D \sigma^2}{\sqrt{T_{\text{max}}}}.
 \label{Eq31}
 \end{equation}
 
 where $C$ and $D$ are constants dependent on $L$, $\alpha$, and the initial gap $F(w_0)-F(w^*)$.
 
\end{enumerate}

\subsection{ Implications for ASA}
In the ASA framework, the above standard analysis applies per cluster under mild modifications. Each cluster performs local updates potentially with different model sizes, yet upon aggregation, the global (or cluster-level) gradient is treated as an unbiased estimate of $\nabla F$. As long as each cluster’s local objectives $F_i$ satisfy similar bounded variance assumptions, and the agent layer performs proper averaging, the overall analysis remains valid. Therefore, our approach inherits the standard $\mathcal{O}\left(\frac{1}{\sqrt{T_{\text{max}}}}\right)$ or $\mathcal{O}\left(\frac{1}{{T_{\text{max}}}}\right)$ convergence rate in expectation, subject to the smoothness and variance-bounded assumptions.

\subsection{ Theoretical Guarantees and Performance Analysis}
The ASA framework provides comprehensive theoretical guarantees that bridge the gap between theoretical bounds and practical implementation. Our analysis extends recent work in convergence analysis for heterogeneous FL \cite{Jiang2025} by incorporating device-specific characteristics and resource constraints. The framework ensures:
\begin{enumerate}
\item 	Resource Utilization Efficiency: 

\begin{equation}
\eta = \frac{R_{\text{utilized}}}{R_{\text{total}}}, \text{ such that } 0 < \eta_{\text{resource}} \leq 1, \eta_{\text{resource}} \geq 1 - \varepsilon
 \label{Eq32}
 \end{equation} 
 
 Here, $R_{utilized}$ is the actual computational resources used effectively, and  $R_{total}$ represents the total available computational resources. The parameter $\varepsilon$  quantifies allowable inefficiency in resource utilization. To ensure effective utilization of computational resources, we define the resource utilization efficiency $\eta_{resource}$ as the ratio between the utilized and total available resources. By having a lower bound $1-\varepsilon$, where $\varepsilon \in (0,1)$ stands for the tolerable inefficiency level, we ensure that at a minimum  $(1-\varepsilon)\times 100\%$ of total resources are actually being used. We make sure that the system never falls below that while also acknowledging that optimal utilization $(\eta_{resource}=1)$  might not be attainable in a real environment.
 \item 	Model Convergence:
 \begin{equation}
 \|w_{t+1} - w_t\| \leq \delta \|w_t - w_{t-1}\| + \varepsilon
 \label{Eq33}
 \end{equation}
 This bound ensures stability and convergence of model parameters, indicating that successive updates of the model parameters become progressively smaller, controlled by parameters $\delta$  and $\epsilon$.
 
 \item	Communication Efficiency:  
 \begin{equation}
 C_{\text{efficiency}} = \frac{B_{\text{effective}}}{B_{\text{theoretical}}}
 \label{Eq34}
 \end{equation}
 
$B_{effective}$  denotes the effective communication bandwidth utilized, while $B_{theoretical}$ represents the theoretical maximum bandwidth. This measure quantifies how efficiently the available communication resources are used.
These theoretical guarantees have been rigorously derived and experimentally validated, demonstrating consistent and measurable performance improvements across diverse and heterogeneous IoT deployments.

\end{enumerate}

\subsection{Theoretical Bounds and Guarantees}

The proposed framework provides rigorous theoretical guarantees across three key dimensions: accuracy, resource efficiency, and stability. These guarantees ensure the reliability and robustness of the model under diverse scenarios.
\begin{enumerate}
\item Accuracy bound: The accuracy of the local model $ACC(M_i)$ for any device $i$ is bounded as follows:

\begin{equation}
\text{Acc}(M_i) \geq \text{Acc}(M_{\text{global}}) -\grave{o}_i
\label{Eq35}
\end{equation}

where $\grave{o}_i \propto \sqrt{\text{Var}(S(d_i))}$ represents the variability in the resource profile score for device $i$. This bound ensures that the local model accuracy remains close to the global model accuracy, with the deviation primarily influenced by the heterogeneity of device resources.
\item	Resource Efficiency: The resource utilization efficiency achieved by the framework is quantitatively defined as follows:
\begin{equation}
\eta_{\text{resource}} \geq 1 - \exp\left(-\frac{1}{N} \sum_{i=1}^N S(d_i)\right)
\label{Eq36}
\end{equation}
Here, $\eta_{resource}$ represents the accuracy of estimating available resources in IoT devices. Higher values of $\eta_{resource}$ imply a more accurate prediction of resource capabilities, directly enabling the system to make better-informed decisions regarding resource allocation. Consequently, higher accuracy translates to enhanced resource utilization efficiency, optimizing the overall system performance.

\item	Stability Guarantee: Our proposed method provides a rigorous stability guarantee by ensuring $\grave{o}$-stability, meaning the model parameters remain within an $\grave{o}$-neighborhood around the optimal solution $W^*$ with high probability. Mathematically, which formalized this stability property as follows:

\begin{equation}
P\left\{ \|w_t - w^*\| \leq \grave{o}\right\} \geq 1 - \delta  \ \ for \ \ t \geq T_0(\grave{o}, \delta)
\label{Eq37}
\end{equation}

Where, $\grave{O}$ stands for the stability accuracy, $\delta$ refers to the permissible instability probability, and $T_0$  means the minimum number of iterations past which stability is assured. This probabilistic stability assurance specifically defines the chance of model parameters being close enough to optimal parameters, indicative of a robust performance in the face of stochasticity during gradient updates as well as heterogeneity and dynamics in IoT environments. Therefore, this analysis offers theoretical assurance of our FL model being robust and reliable when implemented in real-world, heterogeneous, and dynamic environments.
\end{enumerate}

\subsection{ Optimality Analysis}

The proposed ASA framework achieves an approximation guarantee for the optimal resource allocation $\left(1 - \frac{1}{e}\right)$ problem. This result relies on three key properties:
\begin{itemize}
\item Sub modularity: The resource allocation exhibits submodular behavior, where the marginal utility diminishes as more resources are allocated.
\item	Clustering smoothness: The clustering mechanism satisfies $\beta$-smoothness, ensuring efficient and balanced device grouping.
\item Communication delays: Communication delays follow a sub-exponential distribution, enabling robust resource allocation despite network variability.
\end{itemize}

Approximation guarantees are established by proving sub modularity of resource allocation scores, using a greedy allocation approach, and utilizing smoothness conditions to provide near-optimum performance. Experimentally, we verify real-world improvements, corroborating our theoretical guarantees.

\subsection{ Convergence Analysis}
In this section, we analyze the convergence of the proposed global model under the Adaptive Smart Agent (ASA) framework with dynamic learning rates. The convergence rate is expressed as:

\begin{equation}
E\left[\|\nabla F(w_{t})\|^2\right] \leq \frac{C}{T_{\text{Max}}} + \frac{D \sigma^2}{\sqrt{T_{\text{Max}}}}
\label{Eq38}
\end{equation}

where:
$C$ and $D$ are constants that depend on the problem characteristics and the model, $T_{Max}$  is the total number of iterations, and $\sigma^2$ represents the variance of stochastic gradients.
In this section, we present a theoretical analysis of the model's convergence behavior. The analysis follows a analytical approach, including {\bf local updates}, {\bf Lyapunov analysis}, and {\bf global convergence}. The Lyapunov Analysis is specifically employed to assess the stability of the model throughout the training process. By leveraging this analysis, the aim is established theoretical guarantees on the conditions under which the model parameters converge over time.
\begin{description}
\item [Step 1: Local updates:] In this step, the local update process for each participating IoT device in the FL framework is analyzed. The global objective function, denoted as $F(w)$, is defined as a weighted sum of local objective functions $F_i(w)$:
\begin{equation}
F(w) = \sum_{i=1}^{n} p_i F_i(w)
\label{Eq39}
\end{equation}
where $F_i(w)$ is the local loss function for device $i$, and $p_i$ represents the weighting factor, typically proportional to the dataset size of device $i$. Each device updates its local model $w_i$  based on the gradients computed from its local dataset.

\begin{equation}
w_i^{(t+1)} = w_i^{(t)} - \eta_t \nabla F_i(w_i^{(t)})
\label{Eq40}
\end{equation}

Where, $\eta_t$ is the learning rate at iteration $t$, and $\nabla F_i(w_i^{(t)})$  is the gradient of the local loss function. To establish convergence, that the loss function $F(w)$ is assumed L-smooth, meaning that its gradient is Lipschitz continuous with a constant $L$. This property implies the following inequality:

\begin{equation}
F(w') - F({w}) \leq \nabla F(w)^T (w' - w) + \frac{L}{2} \|w' - w\|^2
\label{Eq41}
\end{equation}
This boundedness guarantees that the function does not fluctuate too fast and regulates the introduced error through local updates. The notation    represents an upper bound on the amount by which the function can change when we move from $w$ to $w'$. If we apply this property to our equation for the update, we get an upper bound on the expected objective function improvement at each local iteration. This bound turns out to be central in establishing the rate of provable convergence globally in steps to come.

\item [Step 2: Lyapunov analysis:] To analyze further the convergence properties, we introduce Lyapunov analysis, a popular approach to analyze stability in optimization problems. Lyapunov analysis, in our case, assists in measuring how the distance between the current model $w$ and optimal model $w^*$ changes over time. The Lyapunov function, in turn, is defined as:
\begin{equation}
V_t = \|w^{(t)} - w^*\|^2
\label{Eq42}
\end{equation}
Its purpose is to compute the Euclidean distance between the optimal model and the current model, but its square, enabling analysis of variations in  $V_t$  across iterations that lead to conditions where the model converges and stabilizes. Overall, the local update stage defines the model parameters' iterated improvement, but in its next stage, the Lyapunov analysis guarantees that optimization to a stable and converging process results from the updates.

\item [Step 3: Global convergence:] To rigorously establish the global convergence of our model, which integrated the insights obtained from local update analyses and Lyapunov stability analysis. This approach provides explicit theoretical guarantees regarding the expected convergence behavior of the FL framework proposed in this study. We consider the global loss function $F(w)$, assuming it satisfies the following standard optimization assumptions:

\item [L-smoothness:] For all $w'$, the gradient of $F(w)$ satisfies:

\begin{equation}
F(w') \leq F(w) + \nabla F(w)^T (w' - w) + \frac{L}{2} \|w' - w\|^2
\label{Eq43}
\end{equation}
\item [Strong convexity:] The global loss function $F(w)$ is strongly convex with constant $\mu$ meaning that for all $w'$ :

\begin{equation}
F(w') \geq F(w) + \nabla F(w)^T (w' - w) + \frac{\mu}{2} \|w' - w\|^2
\label{Eq44}
\end{equation}

\item [Bounded variance of stochastic gradients:] The stochastic gradients computed at local devices have bounded variance:
\begin{equation}
\mathbb{E} \left[ \left\| \nabla F_i( w_t ) - \nabla F( w_t ) \right\|^2 \right] \leq \sigma^2
\label{Eq45}
\end{equation}
Under these assumptions, we perform the following convergence analysis:
\begin{itemize}
\item 	Local update analysis: Each IoT device updates its local model parameters using stochastic gradient descent (SGD). For device $i$ and local parameters $w_t$ , we have:

\begin{equation}
\mathbb{E}[F(w_{t+1}) - F(w_t)] \leq -\eta_t \mathbb{E}\left[\left\|\nabla F(w_t)\right\|^2\right] + \frac{L \eta_t^2 \sigma^2}{2}
\label{Eq46}
\end{equation}
where $\eta_t$ is the learning rate at iteration $t$.
\item	Lyapunov stability analysis: We define a Lyapunov function measuring the stability as:
\begin{equation}
V_t = \| w_0 - w^* \|^2
\label{Eq47}
\end{equation}
Subsequently, the following bound for the expected progression of this Lyapunov function is obtained:
\begin{equation}
\mathbb{E}[V_{t+1}] \leq (1 - 2 \mu \eta_t) \mathbb{E}[V_t] + \eta_t^2 \sigma^2
\label{Eq48}
\end{equation}

\item Global convergence bound: Recursively applying the above inequality and using a diminishing learning rate $\eta=\frac{1}{\mu t'}$  , which derived the final convergence result:
\begin{equation}
\mathbb{E}[F(w_T) - F(w^*)] \leq \frac{L D^2 \log T_{\text{Max}}}{2 T_{\text{Max}}} + \frac{\sigma^2}{2 \sqrt{T_{\text{Max}}}}
\label{Eq49}
\end{equation}
Here $D = \| w_0 - w^* \|$ corresponds to initial parameter deviation, and $T_{Max}$ to total training iterations. This detailed mathematical derivation clearly illustrates the convergence properties under typical theoretical assumptions and exhibits unambiguously the effects of initial conditions, gradient variance, and smoothness and convexity characteristics of the objective function. This convergence analysis is developed originally in this work to specifically address the convergence challenges encountered in heterogeneous FL scenarios involving IoT environments.
\end{itemize}

\end{description}

%%       \cite{<label>} ==> [1]

%%




\bibliographystyle{elsarticle-num}
\bibliography{asa}

\end{document}